\documentclass{article}
\usepackage{arxiv}
\usepackage{graphicx} 
\usepackage{hyperref}   
\hypersetup{
	colorlinks=true,
	linkcolor=blue,
	citecolor=blue,
	urlcolor=blue,
}

\usepackage{amsmath,amsfonts}
\usepackage{algorithmic}
\usepackage{array}
\usepackage[caption=false,font=normalsize,labelfont=sf,textfont=sf]{subfig}
\usepackage{textcomp}
\usepackage{stfloats}
\usepackage{url}
\usepackage{verbatim}
\usepackage{graphicx}
\usepackage{xcolor}
\usepackage{placeins}
\usepackage{float}
\def\BibTeX{{\rm B\kern-.05em{\sc i\kern-.025em b}\kern-.08em
    T\kern-.1667em\lower.7ex\hbox{E}\kern-.125emX}}
\usepackage{balance}
\usepackage{hyperref}
\usepackage[vlined,linesnumberedhidden,ruled,resetcount]{algorithm2e}

\title{Real-Time Thermal State Estimation and Forecasting  in Laser Powder Bed Fusion}
\author{Yukta Pareek\\
Department of Mechanical Engineering\\ The Pennsylvania State University\\ University Park, Pennsylvania 16802, USA\\E-mail: ybp5153@psu.edu.\\	
\And
Abdul Malik Al Mardhouf Al Saadi\\
Department of Mechanical Engineering\\ The Pennsylvania State University\\ University Park, Pennsylvania 16802, USA\\E-mail: aaa6554@psu.edu.\\	
\And
Amrita Basak\\
Department of Mechanical Engineering\\ The Pennsylvania State University\\ University Park, Pennsylvania 16802, USA\\E-mail: aub1526@psu.edu.\\	
\And
Satadru Dey\\
Department of Mechanical Engineering\\ The Pennsylvania State University\\ University Park, Pennsylvania 16802, USA\\E-mail: skd5685@psu.edu.\\}

\begin{document}

\maketitle

\begin{abstract}
Laser Powder Bed Fusion (L-PBF) is a widely adopted additive manufacturing process for fabricating complex metallic parts layer by layer. Effective thermal management is essential to ensure part quality and structural integrity, as thermal gradients and residual stresses can lead to defects such as warping and cracking. However, existing experimental or computational techniques lack the ability to forecast future temperature distributions in real time, an essential capability for proactive process control. This paper presents a real-time thermal state forecasting  framework for L-PBF, based on a physics-informed reduced-order thermal model integrated with a Kalman filtering scheme. The proposed approach efficiently captures inter-layer heat transfer dynamics and enables accurate tracking and forecasting of spatial and temporal temperature evolution. Validation across multiple part geometries using measured data demonstrates that the method reliably estimates and forecasts peak temperatures and cooling trends. By enabling predictive thermal control, this framework offers a practical and computationally efficient solution for thermal management in L-PBF, paving the way toward closed-loop control in L-PBF.

\end{abstract}

\section{Introduction and Motivation}
{A}{dditive} {manufacturing (AM) techniques have revolutionized modern manufacturing processes in the energy, medical, aerospace, automotive industries by allowing the production of complex geometries that are often difficult to achieve by traditional methods \cite{DEBROY2018112},\cite{ATTARAN2017677}. Among various AM technologies, laser powder bed fusion (L-PBF) \cite{yadroitsev2021fundamentals} has gained attention due to its ability to create precise, high-performance, yet lightweight parts. In L-PBF, a laser selectively melts and fuses powdered material layer by layer to form a final part. It's layer-by-layer process allows design freedom that traditional subtracting manufacturing cannot match \cite{CHOWDHURY20222109},\cite{NARASIMHARAJU2022375}.

\textcolor{black}{Despite these advantages in L-PBF processes, it faces significant thermal challenges. Uneven temperature distribution can cause warpage, residual stresses, and microstructural inconsistencies, ultimately affecting the mechanical properties of the printed part and result in costly post-processing, inspection, and part rejection \cite{NARASIMHARAJU2022375},\cite{Mazur2022}. Consequently, the quality of the final part is highly dependent on thermal control during the manufacturing process. Accurate control of the thermal state during printing is, therefore, important to ensure part quality and consistency. However, real-time control \cite{mani2017review, mani2017measurement} remain challenging, as highlighted in recent review studies due to a range of technical and computational limitations.}

{Conventional control strategies in L-PBF typically rely on real-time sensor data - such as infrared thermography \cite{grasso2017process} for temperature monitoring or acoustic emissions for melt pool characterization \cite{TAPIA}\cite{CRAEGHS2012753} - collected from the current printing location. While this information informs immediate process control decisions, existing methods generally lack the ability to forecast future thermal behavior. To address this limitation, we present a novel thermal state estimation and forecasting  framework that integrates physics-based modeling with control-theoretic techniques. By formulating the thermal dynamics of the L-PBF process within a state-space representation, our approach enables forecasting of temperature distributions in subsequent layers. This predictive capability supports proactive process control, with the potential to significantly enhance part quality in L-PBF systems.}

\subsection{Literature Review}

The thermal dynamics of the L-PBF process has been widely studied using experimental, physics-based, and data-driven approaches. Existing research has explored various methods for temperature estimation, monitoring, and control, with varying degrees of success in real-time applicability and accuracy.

In L-PBF, real-time temperature measurements can be acquired using embedded thermocouples \cite{hyer2023embedding}, Infrared (IR) pyrometers \cite{mao2023continuous}, or IR cameras \cite{mani2017review}. Thermocouples require direct contact with the part, making their integration challenging due to the need to pause printing, attach the sensor - often by spot welding - and resume the process \cite{mohr2020experimental}. In contrast, IR pyrometers and IR cameras are non-contact sensors that enable continuous thermal monitoring without interfering with the build \cite{gutknecht2020determining}. Pyrometers measure temperature at a single point using emitted infrared radiation \cite{mohr2020experimental}, while IR cameras provide full-field temperature maps \cite{mani2017review}. However, IR cameras present challenges including the need to analyze large volumes of data \cite{clijsters2014situ}, the complexity of maintaining proper focus \cite{lott2011design}, and the difficulty of performing real-time data interpretation \cite{CRAEGHS2012753}.

From a physics-based modeling perspective, in \cite{CRAEGHS2012753}, the authors combined {optical process monitoring} with thermal simulations for defect detection, but their approach was limited to post-process analysis. Similarly, the work in \cite{KING} employed high-fidelity finite element analysis (FEA) to forecast melt pool dimensions and thermal history, but their computational cost made real-time implementation infeasible. {Recent work in \cite{WOOD} addressed a critical gap in controllability and observability of temperature states in L-PBF. Using system-theoretic tools, they formalized the conditions under which thermal states can be estimated from limited sensor data, providing a theoretical foundation for reduced-order modeling. Their results showed that observability depends critically on sensor placement, quantity, and the thermal coupling structure of the system. However, their framework did not address the forecasting of real-time temperature.} The authors in \cite{REN2023299} developed a finite difference method for rapid forecasting and control of part-scale temperature evolution in L-PBF processes. Their model enables fast thermal forecasts, facilitating model-based thermal control strategies. However, it assumes predefined boundary conditions and may require manual tuning for different geometries.

Apart from physics-based modeling, recent advancements in machine learning have enabled data-driven thermal forecasting for L-PBF processes. The work in \cite{SCIME2018114} used convolutional neural networks (CNNs) to correlate thermal images with defect formation, achieving high accuracy but requiring large labeled datasets. Bayesian estimation techniques have also been explored. In \cite{TAPIA}, they applied Gaussian process regression for the forecasting of the melting pool. 
While machine learning models offer computational efficiency by deceasing cost, they often lack physical interpretability. Furthermore, many machine learning algorithms require significant amount of training data to be able to accurately forecast the output.

Hybrid methods combining physics and machine learning have emerged to balance interpretability and adaptability. \cite{SCHEEL} introduced a physics-guided machine learning model (Pulse Approach) for L-PBF thermal analysis, embedding heat transfer principles into neural networks. Their method achieved high accuracy with limited training data but required manual tuning of large number of physical constraints.

While the aforementioned studies have significantly contributed to the advancement of thermal modeling and state estimation in L-PBF processes, several critical limitations persist. Firstly, there exists a fundamental trade-off between computational accuracy and efficiency: high-fidelity FEA methods, although capable of providing precise thermal forecasts, are computationally intensive and therefore unsuitable for real-time process control applications. Conversely, certain machine learning-based approaches, while offering faster inference times, often suffer from limited model interpretability, which impedes their adoption in safety-critical manufacturing environments where understanding the underlying physical phenomena is essential. Secondly, the real-time estimation and predictive forecasting  of the evolving thermal state within the L-PBF process remains an underexplored area, despite its potential to enable proactive process adjustments and improve the quality and consistency of the fabricated parts.

\subsection{Main Contribution and Paper Summary}

In light of the aforementioned limitations, this work introduces a novel framework for real-time thermal state estimation and forecasting  in L-PBF additive manufacturing. The primary objective is to overcome the critical challenge of accurately forecasting the spatiotemporal thermal behavior not only within the currently printed layer but also across future layers - an essential capability for implementing proactive thermal control strategies. 

To this end, we develop a physics-informed, reduced-order thermal model that captures the dominant heat transfer dynamics while remaining computationally efficient. The model is designed to adapt to varying part geometries and layer-wise thermal interactions, making it suitable for complex and evolving build environments. Furthermore, we integrate this reduced-order model with a state estimation scheme based on Kalman filtering techniques, {enabling robust real-time reconstruction and forecasting  of thermal fields from limited and noisy sensor data}. This integrated approach offers a scalable and generalizable solution for predictive thermal management in L-PBF processes, laying the groundwork for advanced closed-loop control architectures aimed at improving part quality.

The remainder of this paper is structured as follows. Section II introduces the mathematical modeling of thermal dynamics in L-PBF processes, Section III details the thermal state estimation and forecasting framework. Section IV presents the simulation results while Section V concludes the paper.

\begin{figure}[H]
\centering
\includegraphics[width=0.4\textwidth]{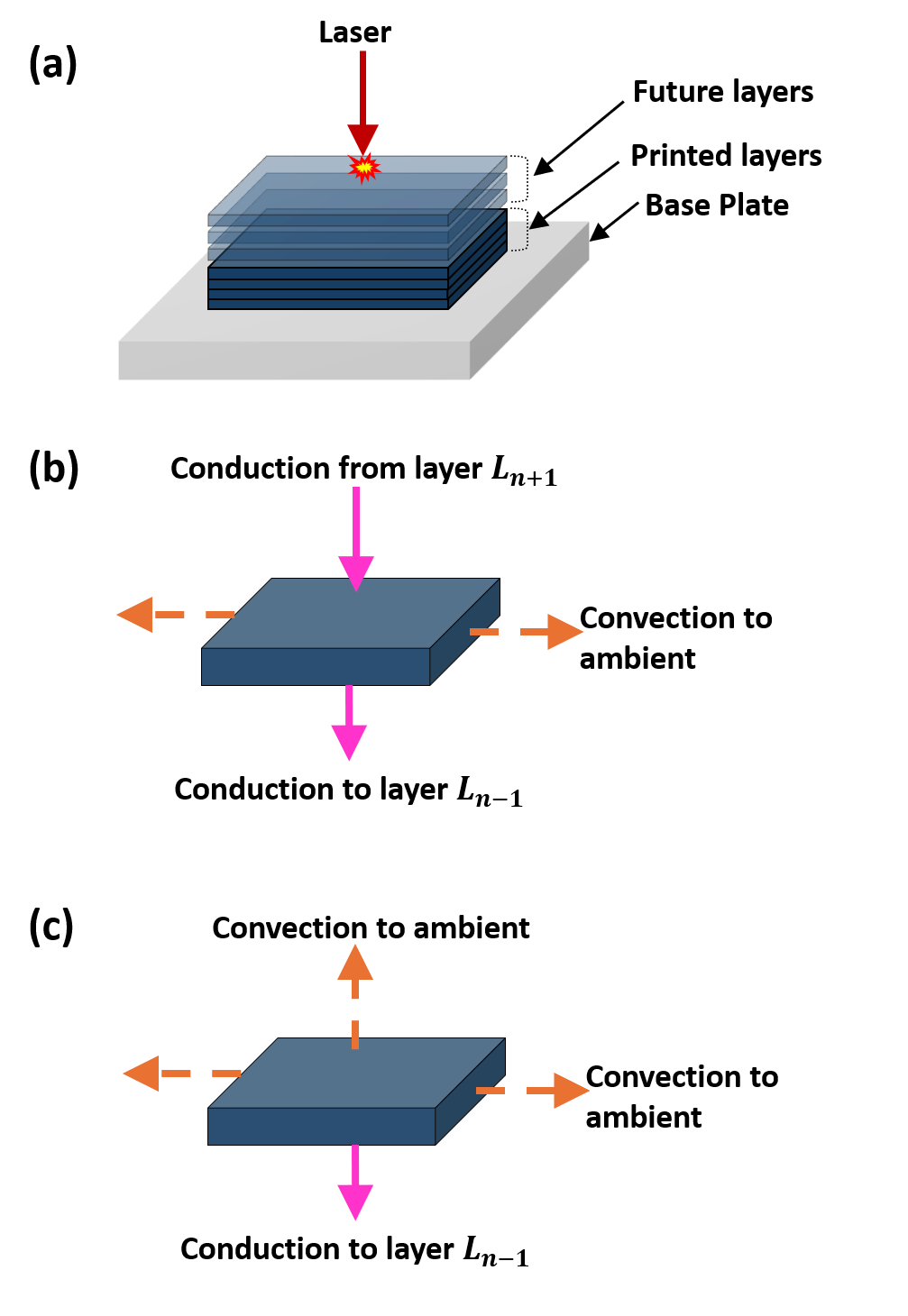}
\caption{(a) A schematic of an example L-PBF part being fabricated in layers. The heat transfer mechanisms for (b) an intermediate layer and (c) the top layer.}
\label{LPBF-Ex}
\end{figure}

\section{Mathematical Modeling of Thermal Dynamics in L-PBF}

A schematic of an example L-PBF part in the process of being manufactured is illustrated in Fig. \ref{LPBF-Ex}, where the printed layers are shown on top of the base plate, along with the future layers. We will use this schematic to explain the modeling approach.

\subsection{Governing Equations for the Thermal Dynamics of a Single Layer}
The thermal behavior for each layer in the L-PBF additive manufacturing (as shown in Fig. \ref{LPBF-Ex}a) is governed by the transient heat transfer phenomena, accounting for phase changes (powder-to-melt-to-solid). In order to capture such phenomena, we resort to a reduced-order ordinary differential equation (ODE) model that captures the essential thermal dynamics while remaining computationally efficient for real-time estimation. Our choice of model is based on the thermal energy balance principle as given in \cite{cengel2015}. In our modeling approach, we assume that the laser application time scale is much smaller than that of the subsequent heat transfer process among the current layer, previous layers, and the environment, after the laser scan is completed. Hence, we treat the high temperature state of the printed layer as the initial temperature for the subsequent heat transfer dynamics. Accordingly, the thermal dynamics of an intermediate layer $n$, depicted in Fig. \ref{LPBF-Ex}b, is given by \cite{cengel2015}: 

\begin{align}
    {\rho} {C_p} {V} \frac{d T_{L_n}}{d t} &= 
    -\underbrace{\frac{K_1}{d}{A_1}(T_{L_n} - T_{L_{n-1}})}_{\substack{\text{Heat conduction} \\ \text{to layer below}}} 
    + \underbrace{\frac{K_2}{d}{A_1}(T_{L_{n+1}} - T_{L_n})}_{\substack{\text{Heat conduction} \\ \text{from layer above}}} \nonumber\\
    &- \underbrace{{\gamma}{A_2}(T_{L_n} - T_{\infty})}_{\substack{\text{Convective heat loss to} \\ \text{ambient via side surfaces}}}, \label{Heat ODE}
\end{align}

The energy balance for layer $n$ accounts for how heat enters and leaves the layer over time. The initial state is given by $T_{L_n}(0)=T_{MP}$ where $T_{MP}$ is the maximum temperature of the printed layer right after the laser scan is completed. The first term represents heat conduction to the layer below. This is treated as a loss because, layer $n$ is warmer than the layer beneath it and, therefore, heat naturally flows downward due to the temperature gradient. The second term represents heat conduction from the layer above. This is a gain, as heat flows into layer $n$ when the layer above it is at a higher temperature. The temperature difference drives this downward heat transfer. The third term accounts for convective heat loss to the ambient environment. The temperature of layer $n$ is higher than the surrounding air and, therefore, heat is lost from the surface to the environment through convection.

In Equation \ref{Heat ODE}, $T_{L_n}, T_{L_{n-1}}, T_{L_{n+1}}$ are the temperatures of layer $n$, layer $(n-1)$ and layer $(n+1)$ in unit $^o C$ respectively; and $T_{\infty}$ = 27 $^o C$ is the temperature of the surrounding environment. $T_{MP}$ is calculated by taking the highest temperature values for a particular layer taken from the measured data. $V$ is the volume of the $n^{th}$ layer, while $A_{1}$ and $A_{2}$ are the total areas of the top/bottom and side surfaces of the $n^{th}$ layer, respectively. $d$ is the layer thickness. $\rho$ is the density, $C_p$ is the specific heat, $K_1$ and $K_2$ are the conductive heat transfer coefficient, and $\gamma$ is the convective heat transfer coefficient. 

In Equation \ref{Heat ODE}, it is assumed that layer $n$ is an intermediate layer. Therefore, the second term on the right-hand side represents heat conduction (Fig. \ref{LPBF-Ex}b). If layer $n$ is the final layer, this term would instead represent heat convection and radiation (Fig. \ref{LPBF-Ex}c). Hence, the updated Equation \ref{Heat ODE} would become:

\begin{align}
    {\rho} {C_p} {V} \frac{d T_{L_n}}{d t} &= 
    -\underbrace{\frac{K_1}{d}{A_1}(T_{L_n} - T_{L_{n-1}})}_{\substack{\text{Heat conduction} \\ \text{to layer below}}} 
    - \underbrace{{\gamma}{A_1}(T_{L_n} - T_{\infty})}_{\substack{\text{Convective heat loss to} \\ \text{ambient via top surface}}} \nonumber\\
    &- \underbrace{{\gamma}{A_2}(T_{L_n} - T_{\infty})}_{\substack{\text{Convective heat loss to} \\ \text{ambient via side surfaces}}}, \label{Heat ODE top}
\end{align}

Equation \ref{Heat ODE top} can be re-written as:

\begin{align}
    {\rho} {C_p} {V} \frac{d T_{L_n}}{d t} 
    &= -\frac{K_1}{d}{A_1}(T_{L_n} - T_{L_{n-1}}) \nonumber\\
    & -{\gamma}{(A_1+A_2)}(T_{L_n} - T_{\infty}), \label{Heat ODE top mod}
\end{align}

Finally, Equation \ref{Heat ODE} can be rewritten by lumping together thermo-physical properties and geometric parameters into constant coefficients, resulting in the following form:

\begin{align}
    {C_1} C_2 \frac{d T_{L_n}}{d t} &= 
   - {C_3}(T_{L_n} - T_{L_{n-1}}) 
   + {C_4}(T_{L_{n+1}} - T_{L_n}) \nonumber\\
    &- {C_5}(T_{L_n} - T_{\infty}), \label{Heat ODE main}
\end{align}

{The parameters $C_{1}$$\big[\frac{Kg}{m^3}\big]$, $C_{2}$$\big[\frac{J}{Kg \cdot K}\big]$, ($C_{3}$, $C_{4}$ and $C_{5})$$\big[\frac{W}{m^3 \cdot K}\big]$ in Equation \ref{Heat ODE main} are the effective parameters (including relevant signs), serving as numerical tuning handles to achieve accurate model fitting with reference data. Rather than strictly enforcing physically exact values for each parameter, the primary objective is to ensure that the reduced-order model output matches the temperature evolution observed in high-fidelity simulations. This approach is justified by the fact that reduced-order models inevitably simplify or neglect certain physical phenomena (e.g. spatial thermal gradients, detailed boundary conditions, phase-change dynamics, geometric factors or time-dependent material properties) present in more comprehensive high-fidelity models. Consequently, the effective parameters compensate for these modeling approximations. This ensures that the reduced order model remains physically consistent and practically suitable for real-time thermal forecasting and control. Equation \ref{Heat ODE main} is intuitive, as it highlights that the temperature evolution of a layer depends on conductive losses to the layer below and convective losses to the environment when it is the top layer. For intermediate layers, conductive input from the layer above also contributes to the heat balance.} 

\subsection{Governing Equations for the Thermal Dynamics of Multiple Layers}
Next, in this subsection, we expand the single layer thermal dynamics presented in Equation \ref{Heat ODE main} to multiple layers of the part under manufacturing. A schematic of this multi-layer modeling framework is shown in Fig. \ref{Method1}. 

\begin{figure*}[h!]
\centering
\includegraphics[width=0.8\textwidth]{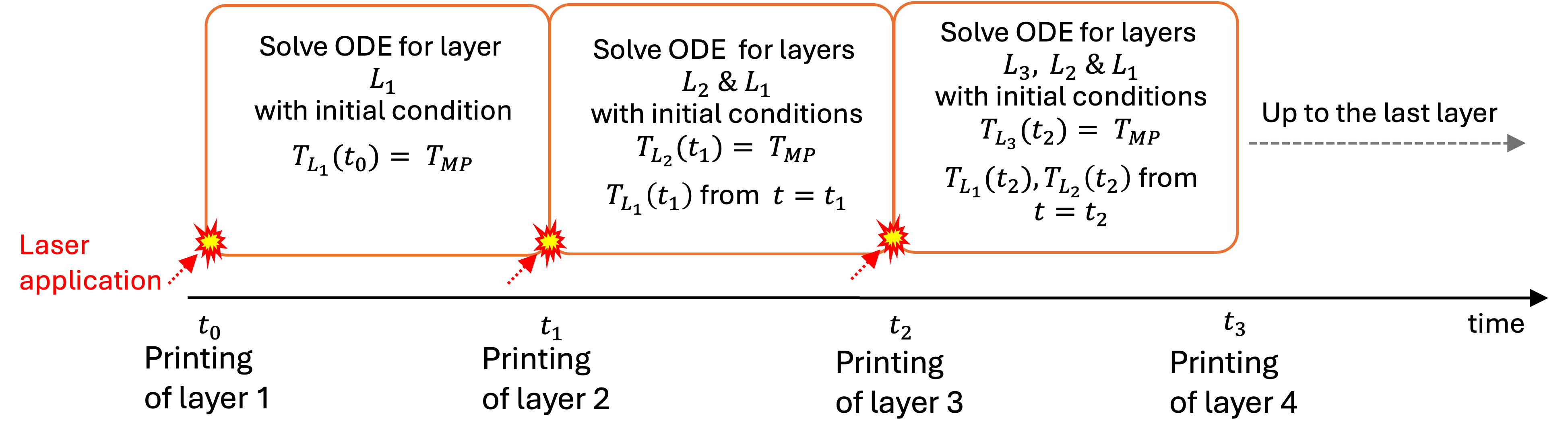}
\caption{A sequential approach for modeling the thermal behavior of multiple layers in L-PBF manufacturing process. Here, $L_i$ denotes layer $i$, laser is applied at $t=t_0$, $t=t_1$, and $t=t_2$ for printing layers $1$, $2$, and $3$, respectively.}
\label{Method1}
\end{figure*}

As shown in Fig. \ref{Method1}, this framework sequentially solves the thermal dynamics as the layers keep being printed. For example, after the first layer is printed and the laser is removed at $t=t_0$, we solve the thermal ODE \eqref{Heat ODE main} for layer $1$ until the application of the laser again for printing layer $2$ at $t=t_1$. That is, we solve the following for $t \in [t_0,t_1]$: 
\begin{align}
    C_1 C_2 \frac{d T_{L_1}}{d t} &=  -C_3(T_{L_1}-T_{B}) \nonumber \\ &+C_4(T_{L_{2}}-T_{L_1})-  C_5(T_{L_1}-T_{\infty}), \label{Heat ODE main 1}
\end{align}
where $T_B$ is the temperature of base plate kept at 27 $^o C$. As indicated earlier, the initial condition is $T_{L_1}(0)=T_{MP}$, which is the highest temperature value taken from the measured data. Then, after the second layer is printed and the laser is removed at $t=t_1$, we solve the thermal ODE \eqref{Heat ODE main} for layers $1$ and $2$ simultaneously until the application of laser again for printing layer $3$ at $t=t_2$. That is, we solve the following for $t \in [t_1,t_2]$: 
\begin{align}
    C_1 C_2 \frac{d T_{L_1}}{d t} &=  -C_3(T_{L_1}-T_{B}) \nonumber \\ &+C_4(T_{L_{2}}-T_{L_1})-  C_5(T_{L_1}-T_{\infty}), \label{Heat ODE main 21}\\
    C_1 C_2 \frac{d T_{L_2}}{d t} &= - C_3(T_{L_2}-T_{L_1}) \nonumber \\ &+C_4(T_{L_{3}}-T_{L_2})-  C_5(T_{L_2}-T_{\infty}), \label{Heat ODE main 22}
\end{align}
where $T_{L_2}(t_1)=T_{MP}$ and $T_{L_1}(t_1)$ is taken from the end point of \eqref{Heat ODE main 1}. Then, after the third layer is printed and the laser is removed at $t=t_2$, we solve the thermal ODE \eqref{Heat ODE main} for layers $1$, $2$ and $3$ simultaneously until the application of laser again for printing layer $4$ at $t=t_3$. That is, we solve the following for $t \in [t_2,t_3]$: 
\begin{align}
    C_1 C_2 \frac{d T_{L_1}}{d t} &= - C_3(T_{L_1}-T_{B}) \nonumber \\ &+C_4(T_{L_{2}}-T_{L_1})-  C_5(T_{L_1}-T_{\infty}), \label{Heat ODE main 31}\\
    C_1 C_2 \frac{d T_{L_2}}{d t} &= -C_3(T_{L_2}-T_{L_1}) \nonumber \\ &+C_4(T_{L_{3}}-T_{L_2})-  C_5(T_{L_2}-T_{\infty}), \label{Heat ODE main 32}\\
     C_1
     C_2 \frac{d T_{L_3}}{d t} &=  -C_3(T_{L_3}-T_{L_2}) \nonumber \\ &+C_4(T_{L_{4}}-T_{L_3})-  C_5(T_{L_3}-T_{\infty}), \label{Heat ODE main 33}
\end{align}
where $T_{L_3}(t_2)=T_{MP}$, and $T_{L_1}(t_2)$ and $T_{L_2}(t_2)$ are taken from the end point of \eqref{Heat ODE main 21}-\eqref{Heat ODE main 22}. This process is followed until the last layer of any part is being printed.

\subsection{State-Space Model Formulation}
To enable integration with the state estimation and forecasting  framework, we reformulate the thermal dynamics model described in the previous subsection, in state-space format. For example, we re-write \eqref{Heat ODE main 31}-\eqref{Heat ODE main 33} as follows:
\begin{align}
    \dot{T}_{L} = &A_cT_{L}+B_cu, \label{State-Space-eqn}
\end{align}
where $T_L = [T_{L_1},T_{L_2},T_{L_3}]^{'}$ is the temperature state vector comprising of temperatures of layers $1$, $2$, and $3$, the input is $u=T_\infty$, and the system matrices are given by:
\begin{align}
    A_c=&
    \begin{bmatrix}
        \beta_c & 0 & 0\\
        0 & \beta_c & 0\\
        0 & 0 & \beta_c
    \end{bmatrix}
    ,\text{ with } \beta_c = -\frac{(-C_3 +C_4-C_5)}{C_1 C_2}, \label{matrix A}\\
     B_c=&
    \begin{bmatrix}
        C_5 & C_5 & C_5
    \end{bmatrix}^{'}.\label{matrix B}
\end{align}
Note that the model parameters $\beta_c$ and $C_5$ may change since the thermal properties of each layer may be affected as new layers are being built on old layers. Considering such time-varying nature of the parameter and subsequently discretizing \eqref{State-Space-eqn} using Euler's method, we get the following discrete-time state-space model:
\begin{align}
    T_{L}(i+1) &= \underbrace{A(i)}_{\text{State transition matrix}} \underbrace{T_{L}(i)}_{\text{Thermal state vector}} + \nonumber\\& \underbrace{B(i)}_{\text{Input matrix}} \underbrace{u(i)}_{\text{Control input}}, \label{State-Space-eqn-d}
\end{align}
where $A = (I+\delta_tA_c)$ and $B = \delta_tB_c$ with $I$ being an identity matrix and $\delta_t$ being the sample time, and $k$ is the time index which is related to the actual time by the following: $t=k\delta_t$ with $i=1,\cdots,\infty$. In the next section, we will use this state-space formulation to describe the proposed thermal state estimation and forecasting framework.

\section{Thermal State Estimation and forecasting  Framework}

In this section, we discuss the thermal state estimation and forecasting  framework for 3D printed parts. A schematic of this framework is shown in Fig. \ref{State-forecast1}. As shown in the schematic, the algorithm utilizes real-time temperature data, along with the state-space model discussed in the previous section. 

Figure \ref{forecast_framework} shows the sequential approach taken for estimating and forecasting  the thermal states. At their core, both estimation and forecasting utilize a Kalman filter operating on the discretized state space model \eqref{State-Space-eqn-d} \cite{lewis2017optimal}. At any given point in time $t$, the framework performs two functions: (i) \textit{Estimation} where the thermal states at time $t$ are estimated, given the available measurement at time $t$. In this case, the Kalman filter utilizes a pre-identified state-space model of the past and present layers along with the available temperature measurement at time $t$ as the feedback signal. (ii) \textit{Forecasting} where the future thermal states at time $t+\Delta t$ are predicted, given the available data until time $t$. Here, the Kalman filter utilizes a pre-identified state-space model of the future layers. However, since the future measurements are not available yet, the Kalman filter leverages synthetic pseudo-data, derived from past historical data, as the feedback signal. For both cases, the Kalman filter assumes the following stochastic version \cite{lewis2017optimal} of the state-space model \eqref{State-Space-eqn-d}:

\begin{align}
    {T}_{L}(i+1) &= A(i) T_{L}(i)+B(i)u(i) + \theta_p(i), \label{State-Space-eqn-d-new}\\
    {y}_{L}(i) &= T_L(i) + \theta_m(i), \label{State-Space-eqn-d-new2}
\end{align}
where ${y}_{L}$ is the feedback signal used for forecasting/estimation; $\theta_p$ and $\theta_m$ are the process and measurement noise vectors, which are modelled by zero mean white Gaussian distribution with variances $\Sigma_p$ and $\Sigma_m$, respectively. Next, we discuss estimation and forecasting approaches in detail.

\begin{figure}[h!]
\centering
\includegraphics[width=0.5\textwidth]{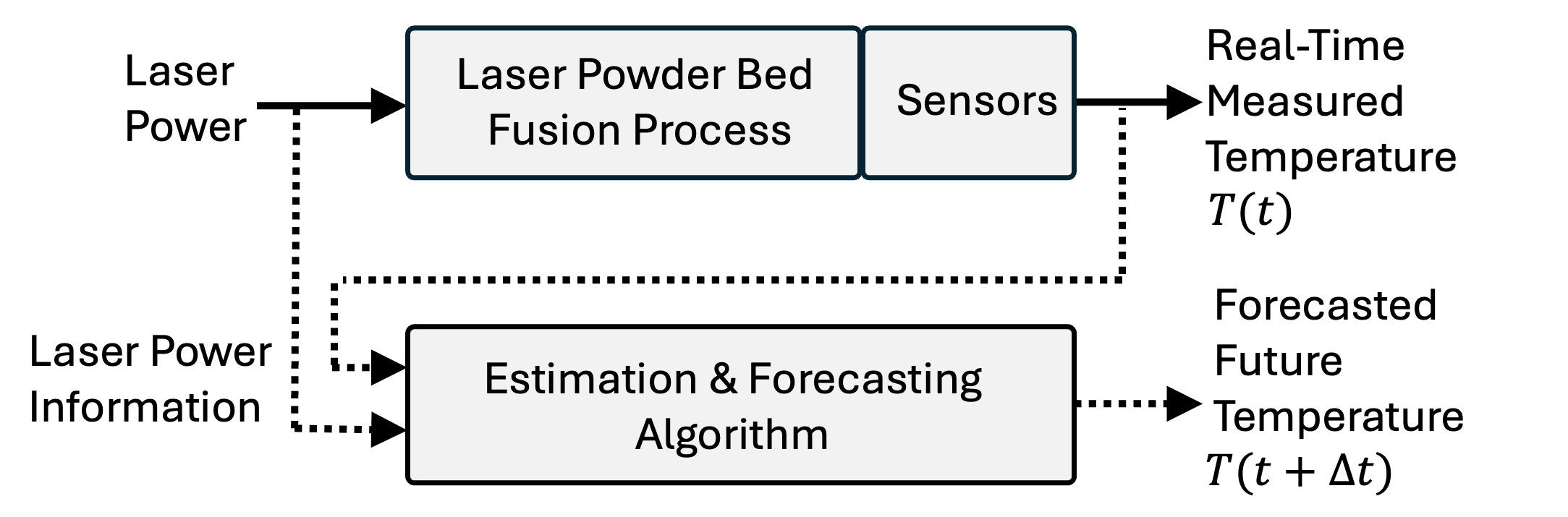}
\caption{A schematic of thermal state estimation and forecasting  framework.}
\label{State-forecast1}
\end{figure}

\begin{figure*}[h!]
\centering
\includegraphics[width=0.8\textwidth]{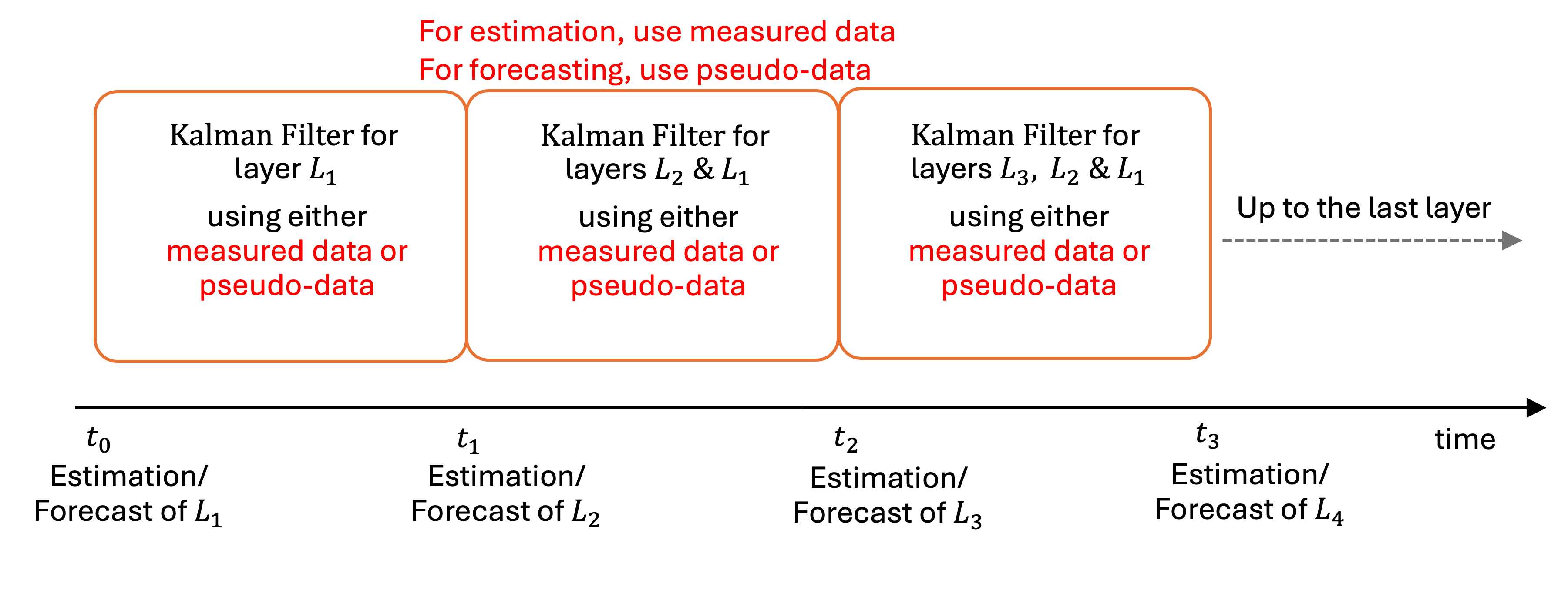}
\caption{A sequential approach for estimating and forecasting  the thermal behavior of multiple layers in L-PBF manufacturing process.}
\label{forecast_framework}
\end{figure*}

\subsection{Thermal State Estimation Framework}
Here, we discuss the formulation of the thermal state estimation problem using the Kalman filtering approach given in \cite{lewis2017optimal}. The estimation problem boils down to computing the thermal state $T_L$ at time index $k$, using the real-time measurement $y_L$ at $k$. To illustrate this, we use the stochastic discrete-time state space model \eqref{State-Space-eqn-d-new}-\eqref{State-Space-eqn-d-new2}. Estimation follows two steps: state propagation and state correction. The estimation cycle begins with the state propagation step, where temperatures are advanced using the open-loop model \eqref{State-Space-eqn-d-new}:
\begin{align}
    \hat{T}_L(i+1)&=A(i) \hat{T}(i) + B(i)u(i), \label{Kalman1}\\
    \hat{P}(i+1)&=A(i){P}(i)A^{'}(i) + \Sigma_p, \label{Kalman2}
\end{align}
where $\hat{T}_L(k)$ represents the propagated open-loop state estimate and $\hat{P}(k+1)$ is the propagated state covariance, given the previous covariance $P(k)$. Next step in estimation is state update using the feedback $y_L$. This update is performed as:
\begin{align}
    K(i)&=P(i)(P(i)+\Sigma_m)^{-1}\label{Kalman3},\\
    \hat{T}_L(i+1)&=\hat{T}_L(i) + K(i)(y_L(i)-\hat{T}_L(i)), \label{Kalman4}\\
    P(i+1) &= (I-K(i))\hat{P}(i+1), \label{Kalman5}
\end{align}
where $K(i)$ is the Kalman gain which is used to combine the open-loop model with the feedback signal, $\hat{T}_L(i+1)$ is the updated state, and $P(i+1)$ is the updated state covariance. The final outcome of the estimation is $\hat{T}_L(i+1)$.

\subsection{Thermal State Forecasting  Framework}
Here, we discuss the formulation of the thermal state forecasting problem using the approach given in \cite{ojeda2013adaptive}. The forecasting problem boils down to computing the thermal state $T_L$ at a future time. However, unlike estimation, we do not have access to the actual measurement $y_L$ at future time instants. In order to address this issue, we utilize \textit{pseudo-data} that would potentially represent the future thermal behavior. Next, we discuss the creation of such \textit{pseudo-data}.

For the first layer, there are no historical data for the part being printed. Hence, for the first layer, we resort to historical first layer data from previous printed similar parts (for example, previously printed parts with similar shape with different dimensions). Based on such historical data, we create a function that describes the shape of the temperature as a function of time. That is: $\tilde{y}_L (t) = f(\lambda,t)$ where $\tilde{y}_L (t)$ is the pseudo-data at a future time $t$ which is represented by the function $f(.)$. The form of this function $f(.)$ can be determined by looking at the time response of the temperature in historical data. The parameters $\lambda$ are adjustable parameters of the function that can be tuned to fit $\tilde{y}_L$ to historical data. For example, minimizing the $\mathcal{L}_2$ vector norm of the error between historical data $y_{historical}$ and the function output, that is, $\min_{\lambda} \left\| y_{historical}(t)-f(\lambda,t)\right\|$ would produce reasonable values of $\lambda$.

For the subsequent layers (printed after the first layer), we use the first layer's data to create pseudo-data. Specifically, pseudo-data $\tilde{y}_L(t)$ are obtained by averaging the historical data of the previous layers. For example, we compute the pseudo-data for the $n$-th layer as
\begin{align}
y_{L_n} = \frac{1}{n-1}\sum_{j=1}^{n-1} T_{L_j}(t\in[t_{j-1},t_{j}]),
\end{align}
where $T_{L_j}(t\in[t_{j-1},t_{j}])$ is the $j$-th layer's temperature data when the $j$-th layer is being built.

Subsequently, as shown in Fig \ref{forecast_framework}, we use the Kalman filtering technique \cite{lewis2017optimal} as in \eqref{Kalman1}-\eqref{Kalman5}; with the following modification: instead of using actual measurement $y_L$ as feedback for state update in equation \eqref{Kalman4}, pseudo-data $\tilde{y}_{L}$ is used as feedback for state update in forecasting step. A summary of the overall algorithmic framework for temperature estimation and forecasting is shown in Algorithm \ref{alg1}. Note that in Algorithm \ref{alg1}, $i$ is the time index that is related to the actual time $t$ as $t=i\delta_t$ with $i=1,\cdots,\infty$.

\begin{algorithm}\label{alg1}
\caption{Thermal State Estimation and Forecasting}
\KwIn{Discrete-time state-space model: $T_L(i+1) = A(i)T_L(i) + B(i)u(i)$. \\
\hspace{1.1cm} {Measured data}: $y_L(i)$. \\
\hspace{1.1cm} {Pseudo-measurement}: $\tilde{y}_L(i)$. \\
\hspace{1.1cm} Noise covariances: $\Sigma_p$, $\Sigma_m$}
\KwOut{Estimated state $\hat{T}_L(i)$; forecasted state $\hat{T}_L(i+\Delta i)$}

\textbf{Initialize:} $\hat{T}_L(0)$, $P(0)$\;

\For{each time step $i$}{
    
    \textbf{forecasting Step (Time Update):} \\
    \Indp
    $\hat{T}_L(i+1) = A(i)\hat{T}_L(i) + B(i)u(i)$ \\
    $\hat{P}(i+1) = A(i)P(i)A(i)' + \Sigma_p$ \\
    \Indm

    \textbf{Determine data type available at time }$i$: \\
    \Indp
    \eIf{$y_L(i)$ is available}{
        $data\_type \leftarrow$ \texttt{"real"}
    }{
        $data\_type \leftarrow$ \texttt{"pseudo"}
    }
    \Indm

    \textbf{Update Step (Measurement Update):} \\
    \Switch{$data\_type$}{
        \Case{\texttt{"real"}}{
            \textbf{(Estimation using real measurement)} \\
            $K(i) = \hat{P}(i+1)(\hat{P}(i+1) + \Sigma_m)^{-1}$ \\
            $\hat{T}_L(i+1) = \hat{T}_L(i+1) + K(i)\left( y_L(i) - \hat{T}_L(i+1) \right)$ \\
            $P(i+1) = (I - K(i))\hat{P}(i+1)$
        }
        \Case{\texttt{"pseudo"}}{
            \textbf{(forecasting using pseudo-measurement)} \\
            $K(i) = \hat{P}(i+1)(\hat{P}(i+1) + \Sigma_m)^{-1}$ \\
            $\hat{T}_L(i+1) = \hat{T}_L(i+1) + K(i)\left( \tilde{y}_L(i) - \hat{T}_L(i+1) \right)$ \\
            $P(i+1) = (I - K(i))\hat{P}(i+1)$
        }
    }
}
\end{algorithm}

\section{Results and Discussion}
{In lieu of experimental measurements, in this paper, we use measured data obtained from a high-fidelity FEA solver to generate spatially and temporally resolved thermal fields, which serve as a surrogate for real-time temperature measurements and provide a controlled environment for developing and validating thermal state estimation algorithms. However, our algorithm is designed to be robust to such uncertainties by incorporating filtering techniques and leveraging temporal correlations in the data to maintain accurate thermal state estimation under realistic operating conditions.}

\subsection{Data Generation for Testing and Validation}
In this section, we discuss the data generation process for testing and validating our reduced order thermal model, discussed in Section II.A and Section II.B. 

\textcolor{black}{Three transient thermal simulations of the L-PBF process were performed with Ansys Mechanical™ using the AM PBF thermal-structural add-on to validate thermal state forecasts. In this setup, a transient thermal model was coupled with a static structural model to forecast both temperature evolution and resulting distortions. For the purpose of this study, only the thermal results were considered for validation. The simulations were conducted on three geometries, each consisting of 10 deposited layers. Each layer has a square cross-section with side lengths of 0.2 mm, 0.4 mm, or 0.8 mm, while the total height of each part is 0.4 mm, corresponding to a uniform layer thickness of 0.04 mm. All parts were simulated on a cubic base with a side length of 2 mm.}

\textcolor{black}{The L-PBF add-on in Ansys Mechanical employs several thermal modeling assumptions to reduce the computational cost of the simulations. These include the use of super layers, layer-by-layer addition, and a power-based heat input method. The super-layers approach groups multiple recently scanned layers into a single simulation layer, based on the assumption that the thermal response of adjacent layers is similar due to the extremely small layer thickness in L-PBF. This reduces the number of elements in the build direction and significantly lowers computation time. The layer-by-layer addition scheme assumes that each layer is introduced and heated uniformly in a single step, neglecting the influence of the laser scanning path on the spatial heat distribution within the layer. Fig. \ref{fig:data_generation_1}(a) illustrates the simulation setup while Fig. \ref{fig:data_generation_1}(b) depicts the mesh. A representative thermal field is shown in Fig. \ref{fig:data_generation_1}(c). Power-based heat input utilizes user-defined laser power and material absorptivity to apply heat across each layer as a whole. The solver internally calculates the heating duration to reflect the time required to scan each layer at a constant scan speed. Applying this method allows the solver to simulate heat application process without the need to specify a scanning pattern.}

\begin{figure}[h!]
\centering
\includegraphics[width=0.4\textwidth]{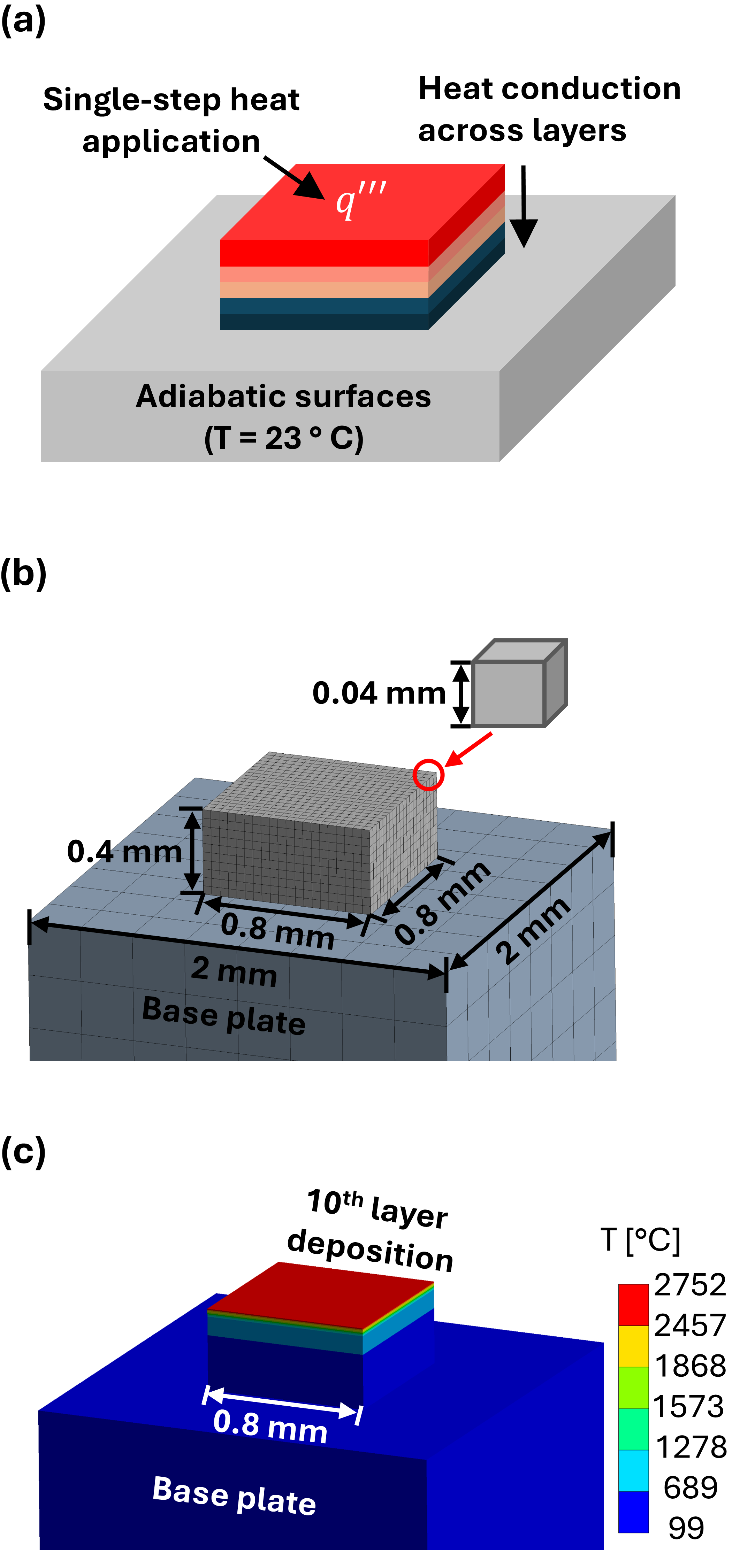}
\caption{Overview of the thermal simulation framework used in Ansys for modeling the L-PBF process. (a) Schematic representation of key thermal assumptions in the AM PBF thermal-structural add-on. (b) Illustrative image showing the Cartesian mesh applied in Ansys. (c) Temperature distribution in [°C] at the timestep corresponding to the deposition of the 10$^{\text{th}}$ layer.}
\label{fig:data_generation_1}
\end{figure}

The building material was assumed to be IN718, the laser power was set to 142 W, the absorptivity to 0.4, and the scan speed to 960 mm/s. These values were selected based on commonly reported processing parameters for Inconel 718 in L-PBF systems  \cite{promoppatum2017comprehensive}, \cite{hwang2022process}. All parts within each simulation were meshed using a Cartesian mesh where the side length of the element of the printed part is 0.04 mm while the side length of the element of the build plate is 0.25 mm. Ansys recommends using one mesh element to represent 10 to 20 layers in order to reduce computational time. However, in this study, only 10 layers were simulated. Therefore, the element size was set equal to the layer height.The simulations were run on Intel Xeon Gold 6230 CPU, featuring a 20-core processor with 40 threads, and equipped with 128 GB RAM. 
Adiabatic boundary conditions were applied to all surfaces, while convection and radiation were neglected. Therefore, the governing heat transfer equation used in the simulation was:

\begin{align}
\rho_h C_{ph} \frac{\partial T}{\partial t} = K_h \left( \frac{\partial^2 T}{\partial x^2} + \frac{\partial^2 T}{\partial y^2} + \frac{\partial^2 T}{\partial z^2} \right) + q''',
\end{align}

{where $\rho_h$ is the material density, $C_{ph}$ is the specific heat capacity, $K_h$ is the thermal conductivity, and $q'''$ is the volumetric heat generation term. The material properties used in the simulations were obtained directly from Ansys Engineering Data, which provides temperature-dependent values in tabular form. These properties reflect the actual physical behavior of the materials used and are independent of the values estimated by the reduced-order model. The volumetric heat generation was computed internally based on user-defined process and geometric parameters such as scan speed, hatch spacing, layer thickness, and superlayer volume. Further details on the modeling framework and underlying assumptions can be found in the Ansys additive manufacturing help documentation \cite{ANSYS}. Throughout the remainder of this paper, the datasets extracted from the analysis conducted in this section are referred to based on the side length of the square base: \textit{Dataset \# 1} corresponds to a 0.2 mm side length, \textit{Dataset \# 2} to 0.8 mm, and \textit{Dataset \# 3} to 0.4 mm.} 

\subsection{Performance of Reduced Order Model}

In this section, we discuss the identification process of the reduced order thermal model discussed in Section II.A and Section II.B. The reduced order model was implemented in MATLAB R2024b on Apple M2 chip, featuring an 8-core CPU and a 10-core GPU, equipped with 8 GB RAM, using Euler's numerical discretization scheme. The parameters of the reduced order model were identified and tested using the training and testing data mentioned in the previous subsection. 

The genetic algorithm (GA) tool, available in the MATLAB Global Optimization Toolbox, was used to fit the parameters. The GA was employed to minimize the root mean square error (RMSE) between the reduced order model’s forecasts and the corresponding ground truth training data. \textit{Dataset \# 1} and \textit{Dataset \# 2}, discussed in the previous subsection, were used as ground truth training data. In the GA setting, the optimization variables were the key parameters including $C_1$, $C_2$, $C_3$, and $C_4$. In other words, the GA performed the following minimization problem:
 \begin{align}
    min_\theta RMSE &= \sqrt{\frac{{\sum ^N_{j=1}}(T_{L_{n_j}}-T_{ANSYS_j})^2}{N}} \label{GA2}
\end{align}
 where $\theta =\begin{bmatrix}
        C_1 &C_2&C_3 & C_4
    \end{bmatrix}$ is the parameter vector to be identified, $T_{L_{n_j}}$ denotes the temperature forecasted by the reduced-order model at the $j$-th layer and $T_{ANSYS_j}$ is the corresponding ANSYS ground truth data. Furthermore, as mentioned in Section II.C, the parameters vary with time, as the effective average thermal properties change every time a new layer is built on top of the old layers. Considering this scenario, the parameters are re-identified every time a new layer is built as shown in Table \ref{table_parameters}. 

\begin{table}[h!]
    \centering
    \caption{Effective parameters values used for layer 1 throughout the printing of all 10 layers.}
    \begin{tabular}{ | m{6em} | m{4em} | m{3em}| m{5em} | m{4em} | }
         \hline
         Layer Number&
          $C_1$& 
         $C_2$ &  
         $C_3$ &  
         $C_4$ \\ 
          \hline
          $L_1$ & 13190 &  0.7&-1500&1500\\ 
          \hline
           $L_2$ & 8190 & 0.7&-5000&5000\\ 
          \hline
           $L_3$ & 8190 & 1.1 &-35000&35000\\ 
           \hline
           $L_4$ & 9190 & 0.9 &-35000&35000\\ 
           \hline
           $L_5$ & 8190 & 1.1 &-40000&40000\\ 
           \hline
           $L_6$ & 8190 & 0.9&-25000&25000\\ 
           \hline
           $L_7$ &  8190& 0.9 &-30000&30000\\ 
           \hline
           $L_8$ &8190  & 0.9&-30000&30000\\ 
           \hline
           $L_9$ & 8190&  0.9&-35000&35000\\ 
           \hline
           $L_{10}$ & 11518.68 & 1.45 &-10463.53&36324.86\\ 
           \hline
    \end{tabular}
    \label{table_parameters}
\end{table}

\begin{figure}[H]
\centering
\includegraphics[width=0.5\textwidth]{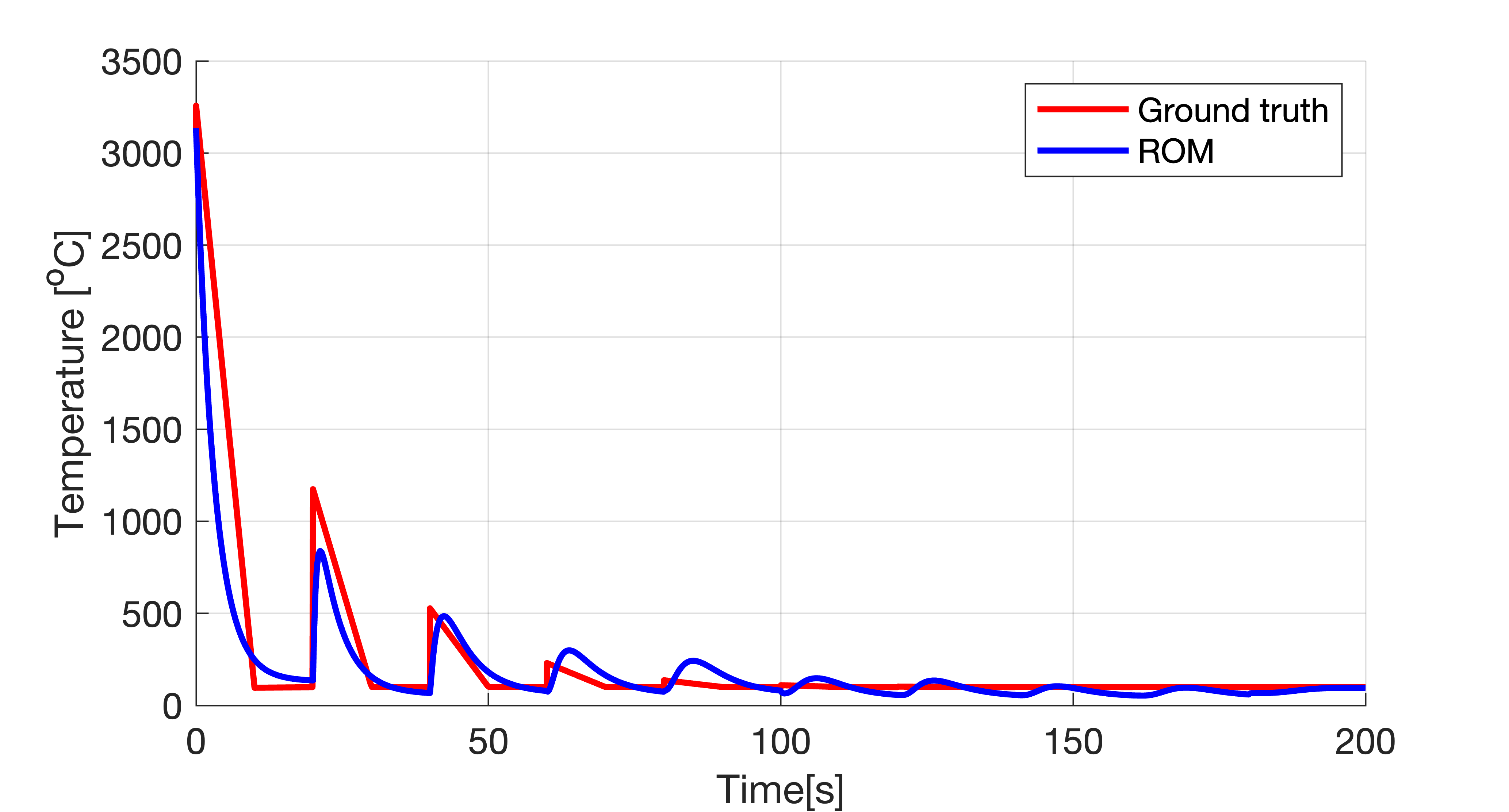}
\caption{Comparison of the reduced order model's forecasting capability (denoted by ``ROM'') and ground truth of Layer 1, for \textit{Dataset \# 1} (cube geometry of size $0.2$ mm).}
\label{ROM1}
\end{figure}

\begin{figure}[H]
\centering
\includegraphics[width=0.5\textwidth]{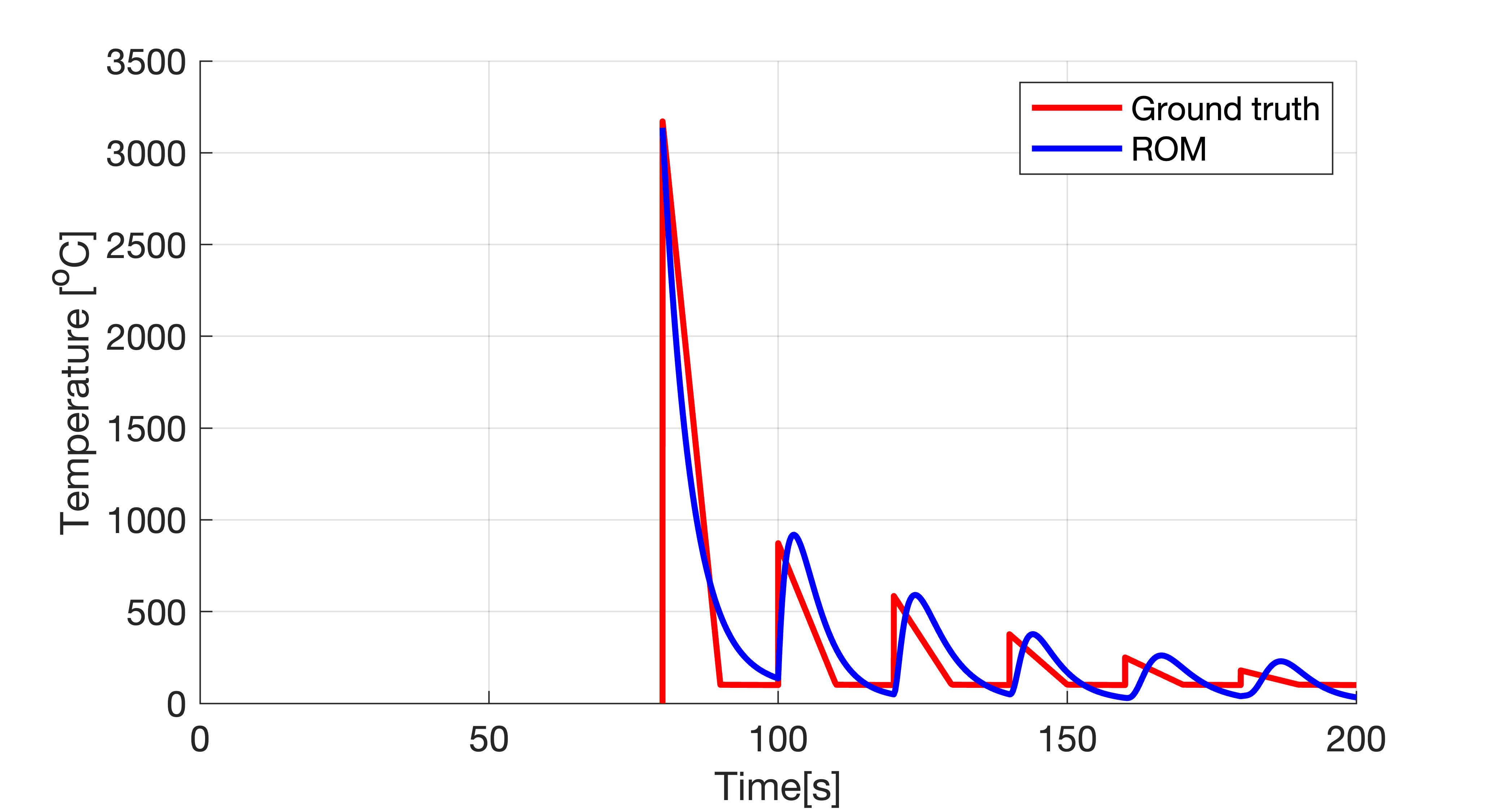}
\caption{Comparison of the reduced order model's forecasting capability (denoted by ''ROM'') and ground truth of Layer 5, for \textit{Dataset \# 3} (cube geometry of size $0.4$ mm)}
\label{ROM2}
\end{figure}

Figures (\ref{ROM1}-\ref{ROM3}) demonstrate the comparative results between the reduced order model and the ground truth data. These figures show the reduced order model's ability to replicate peak temperatures during laser exposure, cooling trends during inter-layer delays, and vertical temperature gradients. These are shown for representative layers: Layer 1 of the $0.2$ mm cube, corresponding to training \textit{Dataset \# 1} (see Fig. \ref{ROM1}); Layer 5 of the $0.4$ mm cube, corresponding to testing Dataset \# 3 (see Fig. \ref{ROM2}); and Layer 10 of a $0.8$ mm cube, corresponding to training \textit{Dataset \# 2} (see Fig. \ref{ROM3}). In all cases, the reduced-order model captures key thermal dynamics with reasonable accuracy. Table \ref{tableROM} shows RMSE values across all 10 layers for the three datasets. The errors gradually reduce for deeper layers, which aligns with increased thermal averaging and stabilization due to material accumulation.

\begin{figure}[H]
\centering
\includegraphics[width=0.5\textwidth]{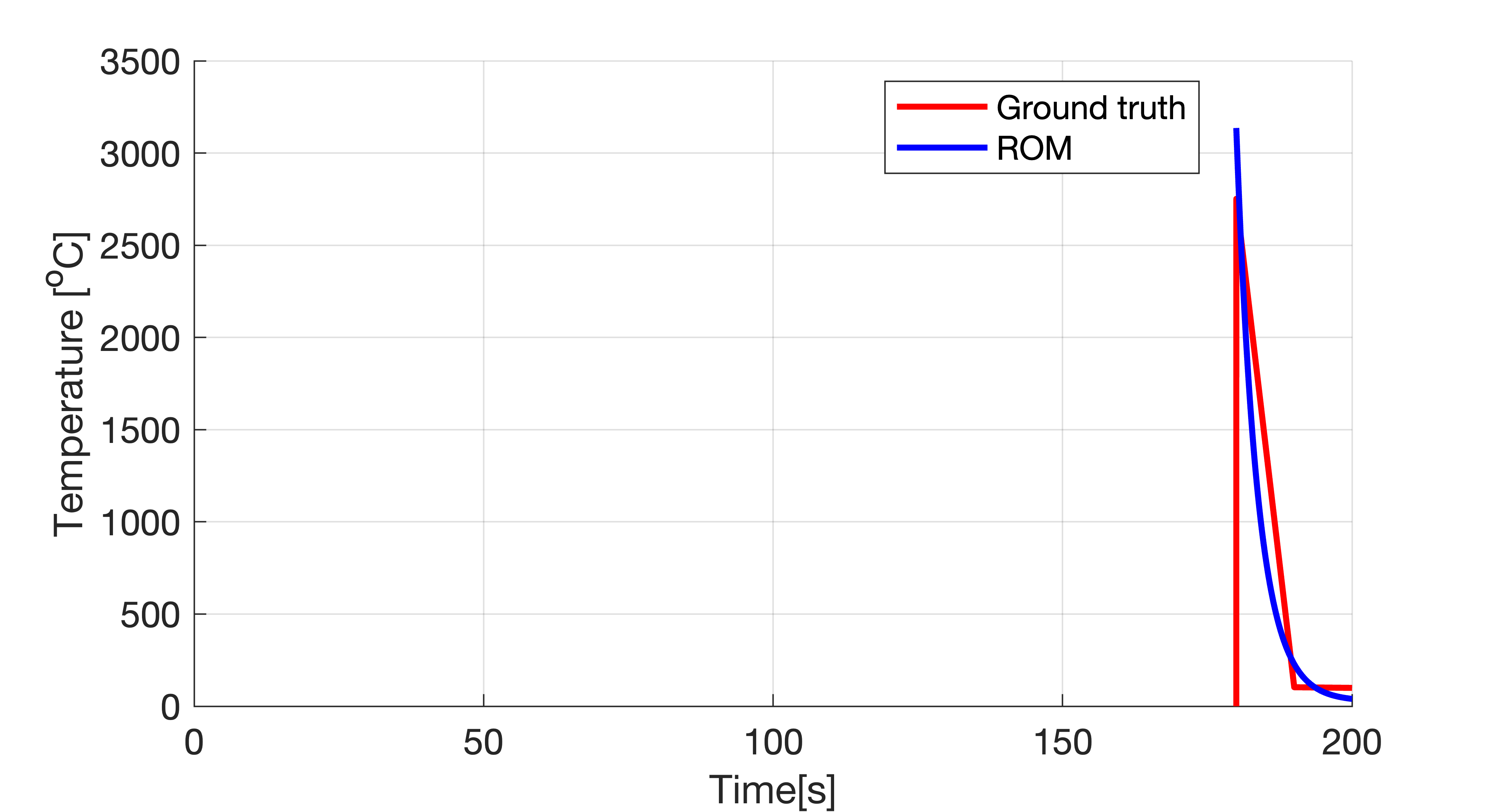}
\caption{Comparison of the reduced order model's forecasting ability (denoted by ''ROM'') and ground truth of Layer 10, for \textit{Dataset \# 2} (cube geometry of size $0.8$ mm).}
\label{ROM3}
\end{figure}

\begin{table}[H]
    \centering
    \caption{RMSE in $^\circ\text{C}$ between the reduced order model's forecasting capability and the ground truth data for all layers of all three datasets.}
    \begin{tabular}{ | m{6em} | m{6em}| m{5em} |m{6em} | }
         \hline
         Layer Number& 
         Training \textit{Dataset \#1} &  
         Testing \textit{Dataset \#3} &  
         Training \textit{Dataset \#2} \\ 
          \hline
          $L_1$ & 181.60 & 162.91 & 128.77 \\ 
          \hline
           $L_2$ & 164.94 & 145.77 &123.05\\ 
          \hline
           $L_3$ & 143.68& 127.18 &120.78\\ 
           \hline
           $L_4$ & 127.64 & 115.76 &117.65\\ 
           \hline
           $L_5$ & 139.28 & 128.39 &123.19\\ 
           \hline
           $L_6$ & 163.69 &  148.64&127.42\\ 
           \hline
           $L_7$ & 131.95 & 119.02 &111.31\\ 
           \hline
           $L_8$ & 130.16  & 117.86 &102.28\\ 
           \hline
           $L_9$ & 114.62 &  86.79&70.13\\ 
           \hline
           $L_{10}$ & 155.42 & 100.20 &97.70\\ 
           \hline
    \end{tabular}
    \label{tableROM}
\end{table}

Next, in Fig. \ref{Interlayer heat transfer coefficient}, we show representative plots of how absolute values of the model parameters such as $C_3$, $C_4$, and $C_2$ vary with time as the new layers are built.

\subsection{Creation of Pseudo-Data for Thermal State forecast}
As discussed in Section III.B, we outline our approach for creating the pseudo-data for thermal state forecasting purposes. From our ground truth data, the function $f(.)$ is found to be a triangle as shown in an example temperature response in Fig. \ref{pseudo-idea-fig}. Accordingly, we used the following function: 
\begin{align}
    \tilde{y}_{L_1}=\frac{(T_p-T_s)}{T_b} + T_p \label{pseudo-data}
\end{align}
where $\tilde{y}_{L_1}$ is the pseduo-data for the first layer, and $T_p, T_s, T_b$ are the average peak temperature, the settling temperature, and the base measurement of the triangle in Fig. \ref{pseudo-idea-fig}, respectively. We found this $\tilde{y}_{L_1}$ by averaging the properties of two triangles from \textit{Dataset \# 1} and \textit{Dataset \# 2}. 

\begin{figure}[h]
\centering
\includegraphics[width=0.5\textwidth]{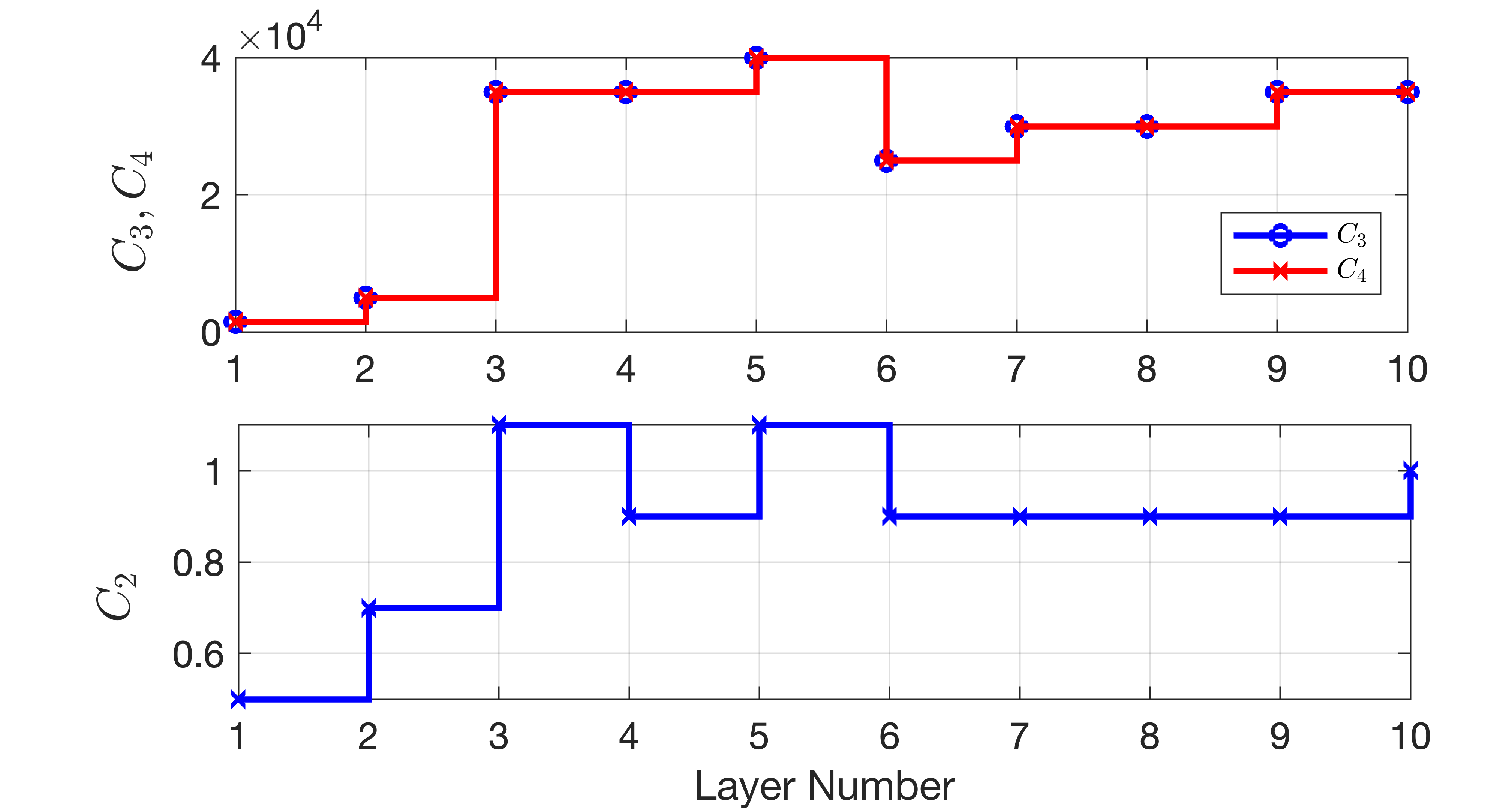}
\caption{Absolute values of the inter-layer model parameters $C_3$, $C_4$, and $C_2$ for Layer 1 of \textit{Dataset \# 3} (cube geometry of size $0.4$ mm).}
\label{Interlayer heat transfer coefficient}
\end{figure}

\begin{figure}[h]
\centering
\includegraphics[width=0.5\textwidth]{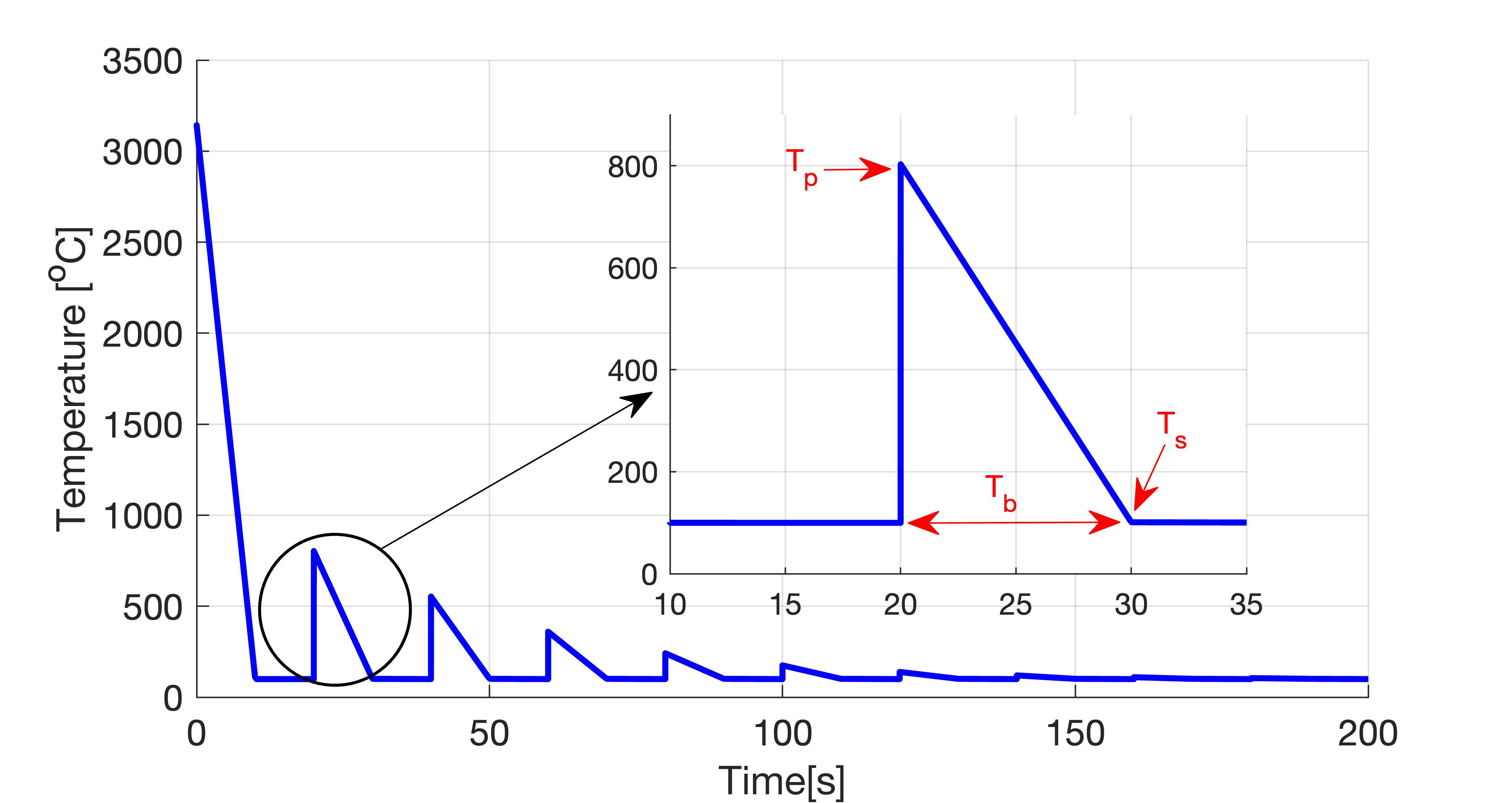}
\caption{Functional form of the historical temperature response to create first layer's pseudo-data.}
\label{pseudo-idea-fig}
\end{figure}

We emphasize that alternative functional forms may also be employed to fit the measured data, thereby enabling broader generalization in the context of thermal state forecasting. While for the current dataset, the triangular representation of the function provided a simple yet effective abstraction of the underlying dynamics (see Fig. \ref{pseudo-idea-fig}), the methodology is not restricted to this choice. Depending on the forecasting objective, system complexity, or computational requirements, other parametric or non-parametric models such as polynomial regressions, exponential functions, or machine learning–based function approximators can be adopted. This flexibility ensures that the pseudo-data generation framework remains adaptable to diverse thermal behaviors and enhances its applicability to a wider range of operational scenarios.

\subsection{Performance of State Estimation and Forecasting  Algorithm}

{The state estimation and forecasting framework is formulated around a Kalman filter applied to a discretized state-space representation of the thermal process. This filter enables real-time reconstruction of the thermal field by iteratively updating the layer temperatures. For the estimation task, the updates are performed using sensor-acquired measurements, whereas for forecasting, pseudo-measurements are synthesized from historical or simulation-derived data to approximate future thermal states. The statistical characterization of the process and measurement uncertainties is encoded through the process noise covariance ($\Sigma_p$) and measurement noise covariance ($\Sigma_m$), respectively. These covariance matrices critically influence the filter’s convergence properties, stability, and overall estimation accuracy. In this study, the covariance values are empirically tuned to achieve reliable performance, with $\Sigma_p = 2.3$ and $\Sigma_m = 1$ selected as representative values.}

Estimation and forecasting were tested on multiple layers of a $0.4$ mm cube geometry (\textit{Dataset \# 3}). For each layer, the Kalman filter was initialized with appropriate prior states and covariance values. We illustrate the results in terms of Layers 1, 4, and 8, which are shown in Figs \ref{Estimation1}-\ref{Estimation3}. In the top plots of these figures, we compare the estimated and forecasted temperatures by the Kalman filter to ground truth (see the top plot of these figures). In the bottom plots of Figs. \ref{Estimation1}-\ref{Estimation3}, the error between the ground truth and state-space model forecasting (denoted as ``open loop model error'') and the error between the ground truth and the Kalman forecasting (denoted as ``Kalman filter error'') are also shown. In each case, both estimation and forecasting follow the thermal trends closely.

\begin{figure}[h]
\centering
\includegraphics[width=0.5\textwidth]{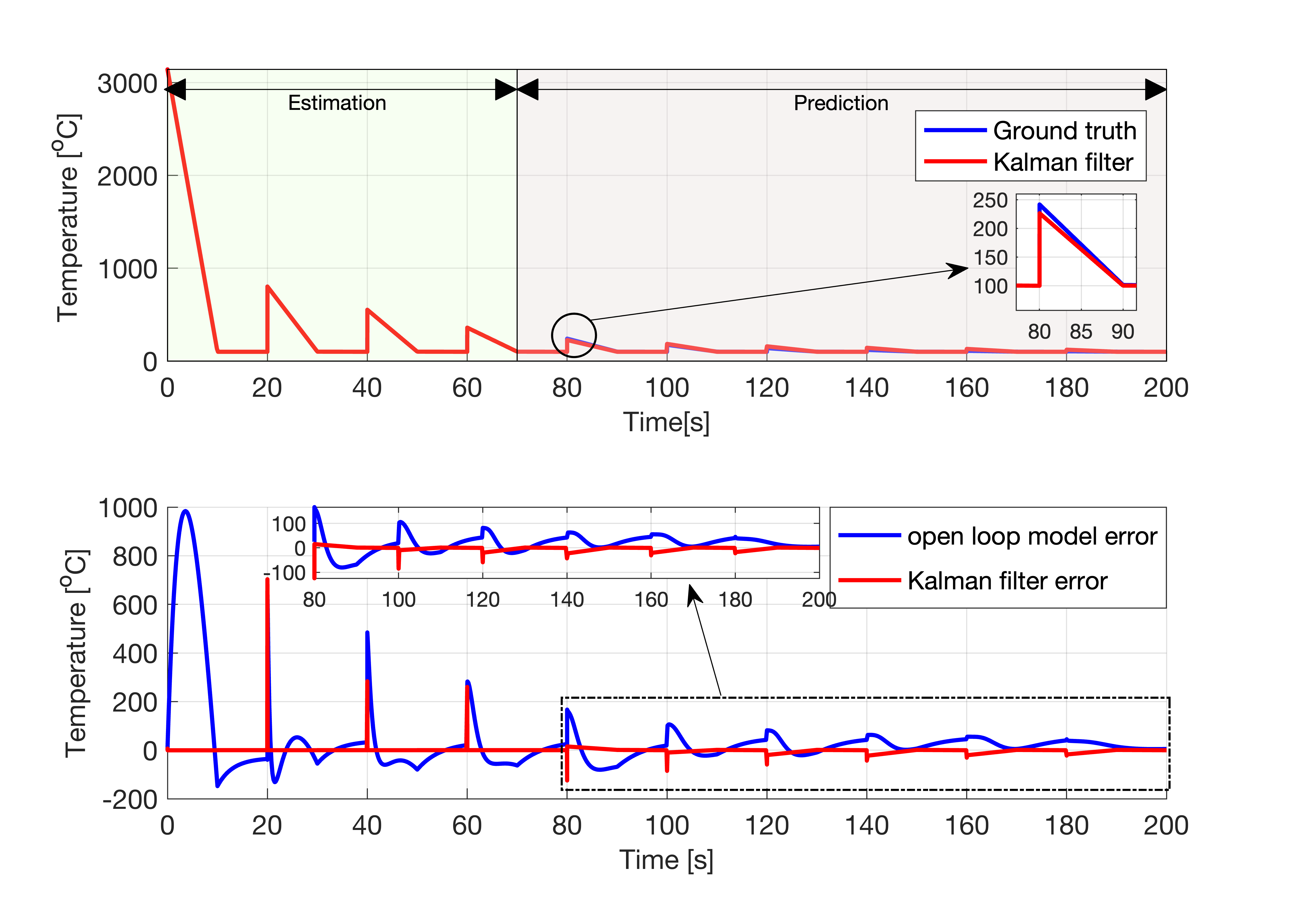}
\caption{\textit{Top Plot:} Estimation and forecasting performance of the proposed Kalman filtering approach for Layer 1 of Dataset \# 3 (cube geometry of size $0.4$ mm). Here, the Kalman filter estimated and forecasted temperatures are compared with the ground truth. The estimation was performed for the period of $t=0-70$ seconds, and thereafter, the forecasting was performed for $t=70-200$ seconds. \textit{Bottom Plot:} Error between the ground truth and state-space model forecasting (denoted as ``open loop model error'') and the error between the ground truth and the Kalman forecasting (denoted as ``Kalman filter error'') for the same specifications.}
\label{Estimation1}
\end{figure}

\begin{figure}[h!]
\centering
\includegraphics[width=0.5\textwidth]{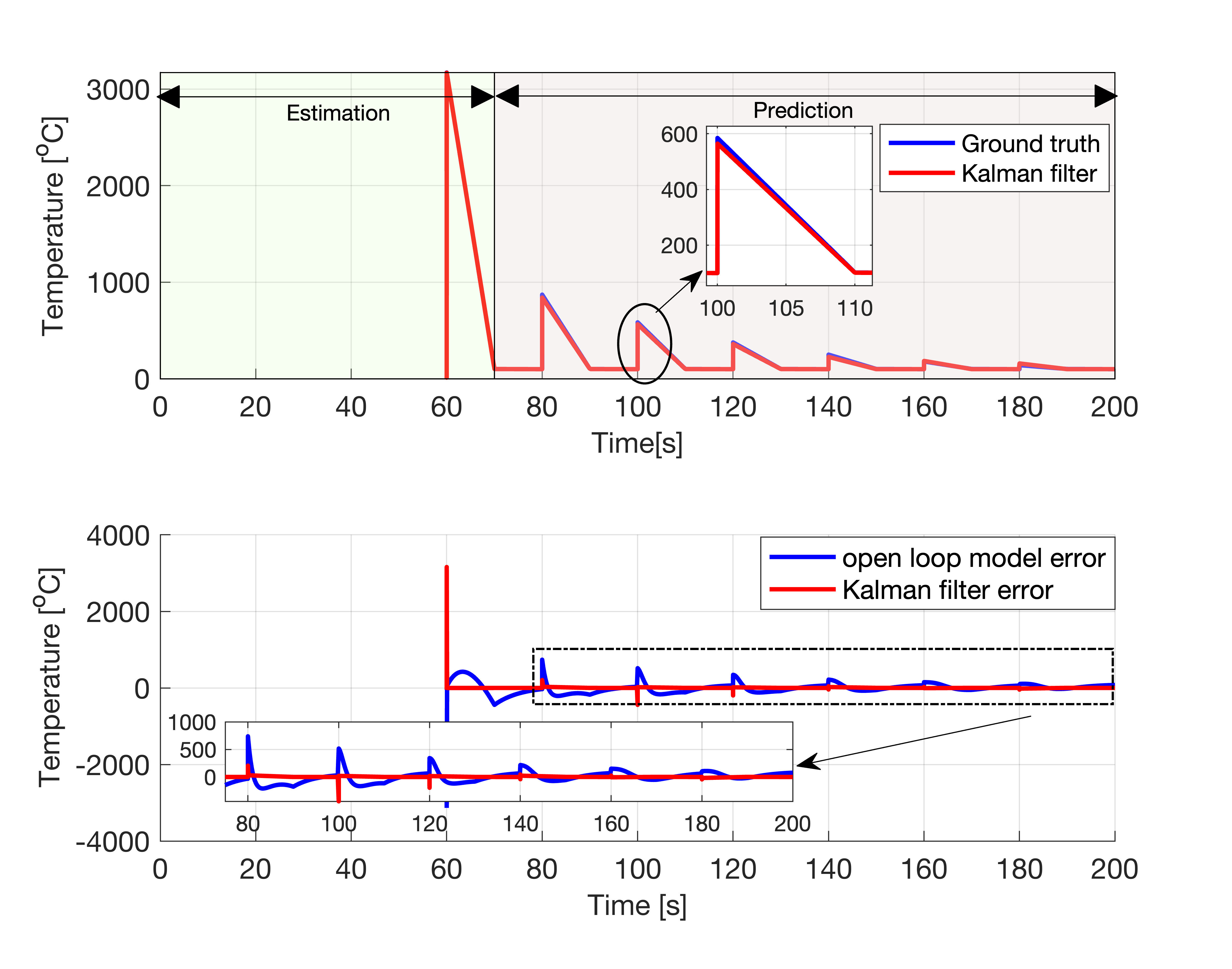}
\caption{Estimation and forecasting performance of the proposed Kalman filtering approach for Layer 4 of \textit{Dataset \# 3} (cube geometry of size $0.4$ mm). Here, the Kalman filter estimated and forecasted temperatures are compared with the ground truth. The estimation was performed for the period of $t=0-70$ seconds, and thereafter, the forecasting was performed for $t=70-200$ seconds. \textit{Bottom Plot:} Error between the ground truth and state-space model forecasting (denoted as ``open loop model error'') and the error between the ground truth and the Kalman forecasting (denoted as ``Kalman filter error'') for the same specifications.}
\label{Estimation2}
\end{figure}

\begin{figure}[h!]
\centering
\includegraphics[width=0.5\textwidth]{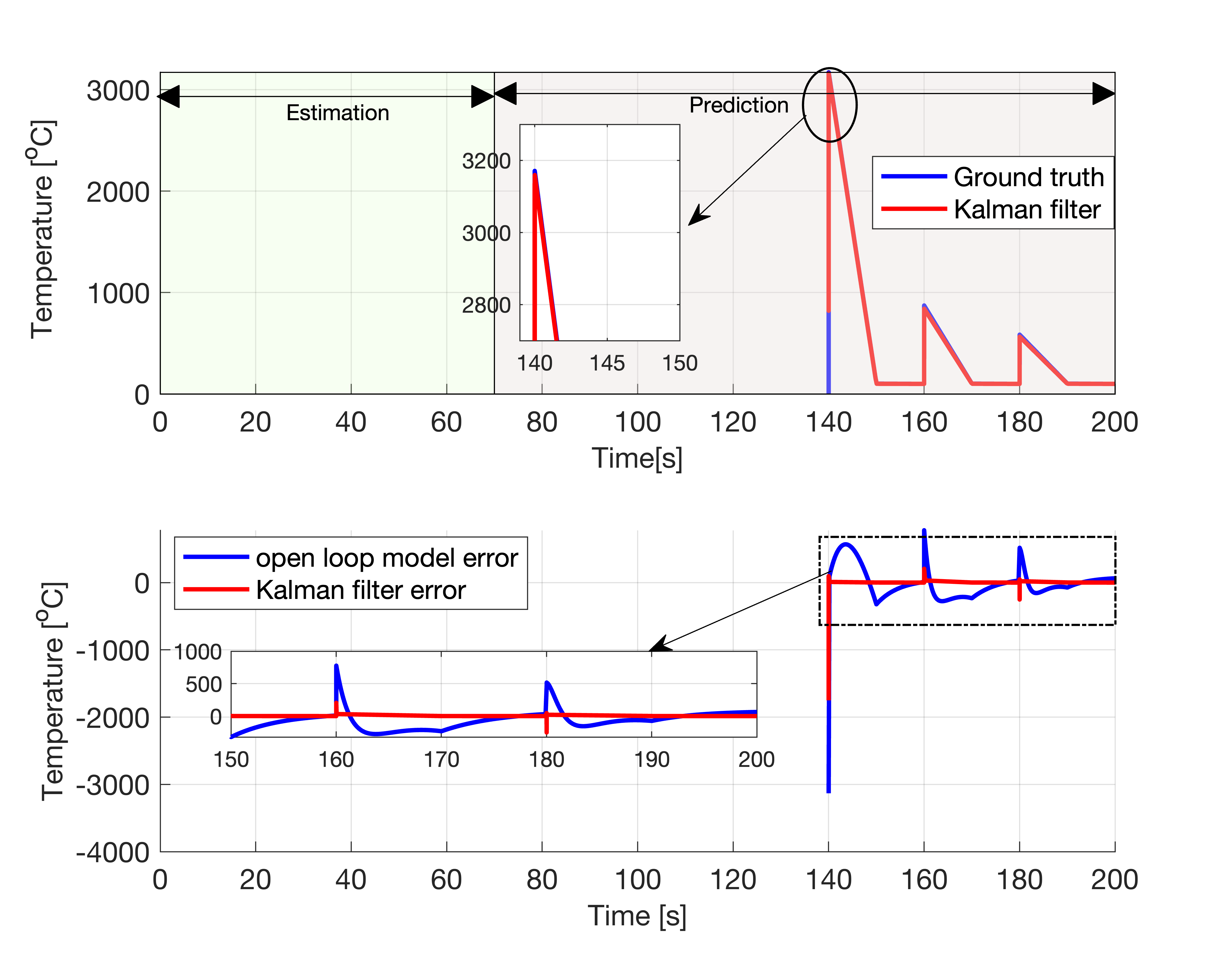}
\caption{Estimation and forecasting performance of the proposed Kalman filtering approach for Layer 8 of \textit{Dataset \# 3} (cube geometry of size $0.4$ mm). Here, the Kalman filter estimated and forecasted temperatures are compared with the ground truth. The estimation was performed for the period of $t=0-70$ seconds, and thereafter, the forecasting was performed for $t=70-200$ seconds. \textit{Bottom Plot:} Error between the ground truth and state-space model forecasting (denoted as ``open loop model error'') and the error between the ground truth and the Kalman forecasting (denoted as ``Kalman filter error'') for the same specifications.}
\label{Estimation3}
\end{figure}

Table \ref{tableKF}, lists the RMSE values of the Kalman filter forecasts across all layers for each geometry. In particular, RMSE decreases consistently with depth of printed part (increased number of layers), particularly in the cases $0.4$ mm and $0.8$ mm. This trend can be attributed to two main factors. First, deeper layers benefit from thermal averaging, as the number of printed layers increases, heat dissipation effects smooth out rapid thermal fluctuations, making the system easier to model. Second, the estimation and forecasting algorithm gains continuity and historical context as more data from earlier layers becomes available, allowing the filter to make increasingly informed forecasts. This cumulative effect results in reduced forecasting error and higher confidence in temperature reconstruction as the build progresses.

\begin{table}[h!]
    \centering
    \caption{RMSE in $^oC$ of Kalman Filter for all layers for their respective geometry scale}
    \begin{tabular}{ | m{6em} | m{5em}| m{5em} | m{5em} | }
         \hline
          Layer Number& 
         Training \textit{Dataset \#1} &  
         Testing \textit{Dataset \#3} &  
         Training \textit{Dataset \#2} \\ 
          \hline
          $L_1$ & 18.54 &  6.03&19.03\\ 
          \hline
           $L_2$ & 54.6 & 24.46&24.28\\ 
          \hline
           $L_3$ & 20.07 & 11.05 &23.04\\ 
           \hline
           $L_4$ & 19.87 & 12.27 &23.31\\ 
           \hline
           $L_5$ & 17.66 & 10.71 &21.49\\ 
           \hline
           $L_6$ & 18.20 & 14.86 &20.33\\ 
           \hline
           $L_7$ &  12.03& 12.67 &13.24\\ 
           \hline
           $L_8$ &9.95  & 10.67 &11.47\\ 
           \hline
           $L_9$ & 8.99 &  9.42&10.06\\ 
           \hline
           $L_{10}$ & 11.03 & 10.68 &10.13\\ 
           \hline
    \end{tabular}
    \label{tableKF}
\end{table}

Fig. (\ref{PK1}-\ref{PK3}) shows the evolution of state covariance and Kalman gain throughout the printing process. The plot shows state covariance and Kalman gain of the proposed Kalman filtering approach with respect to time; for layer 1, 4 and 8 of \textit{Dataset $3$}, respectively. Overall, the results demonstrate that the proposed reduced-order model, when augmented with Kalman filtering, provides a robust and efficient framework for real-time thermal state estimation and forecasting in L-PBF. 

\begin{figure}[H]
\centering
\includegraphics[width=0.5\textwidth]{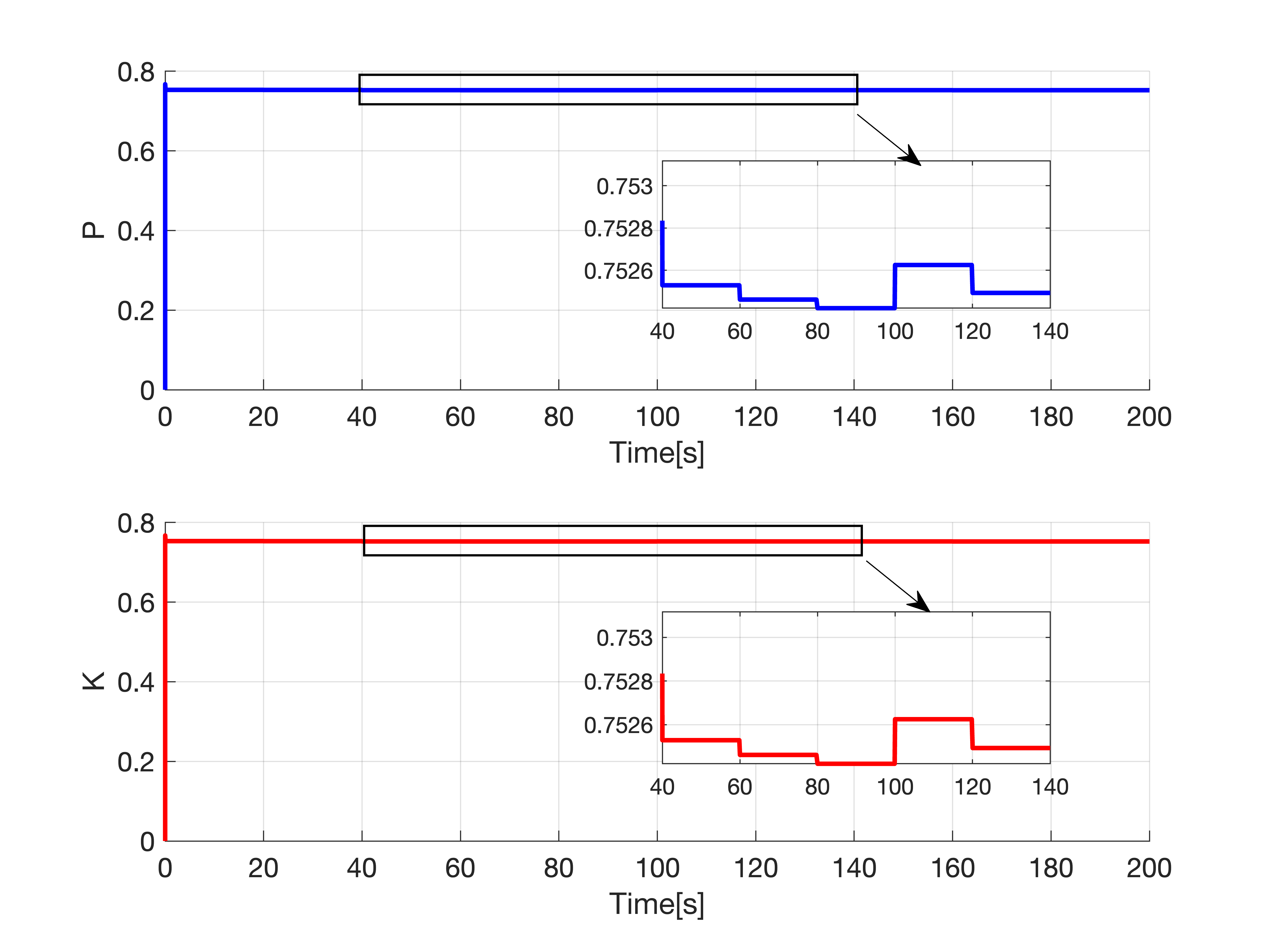}
\caption{\textit{Top Plot:}  Evolution of state covariance of the proposed Kalman filtering approach for Layer 1 of \textit{Dataset \# 3} (cube geometry of size $0.4$ mm). \textit{Bottom Plot:}  Evolution of Kalman gain of the proposed Kalman filtering approach for Layer 1 of \textit{Dataset \# 3} (cube geometry of size $0.4$ mm).}
\label{PK1}
\end{figure}

\begin{figure}[H]
\centering
\includegraphics[width=0.5\textwidth]{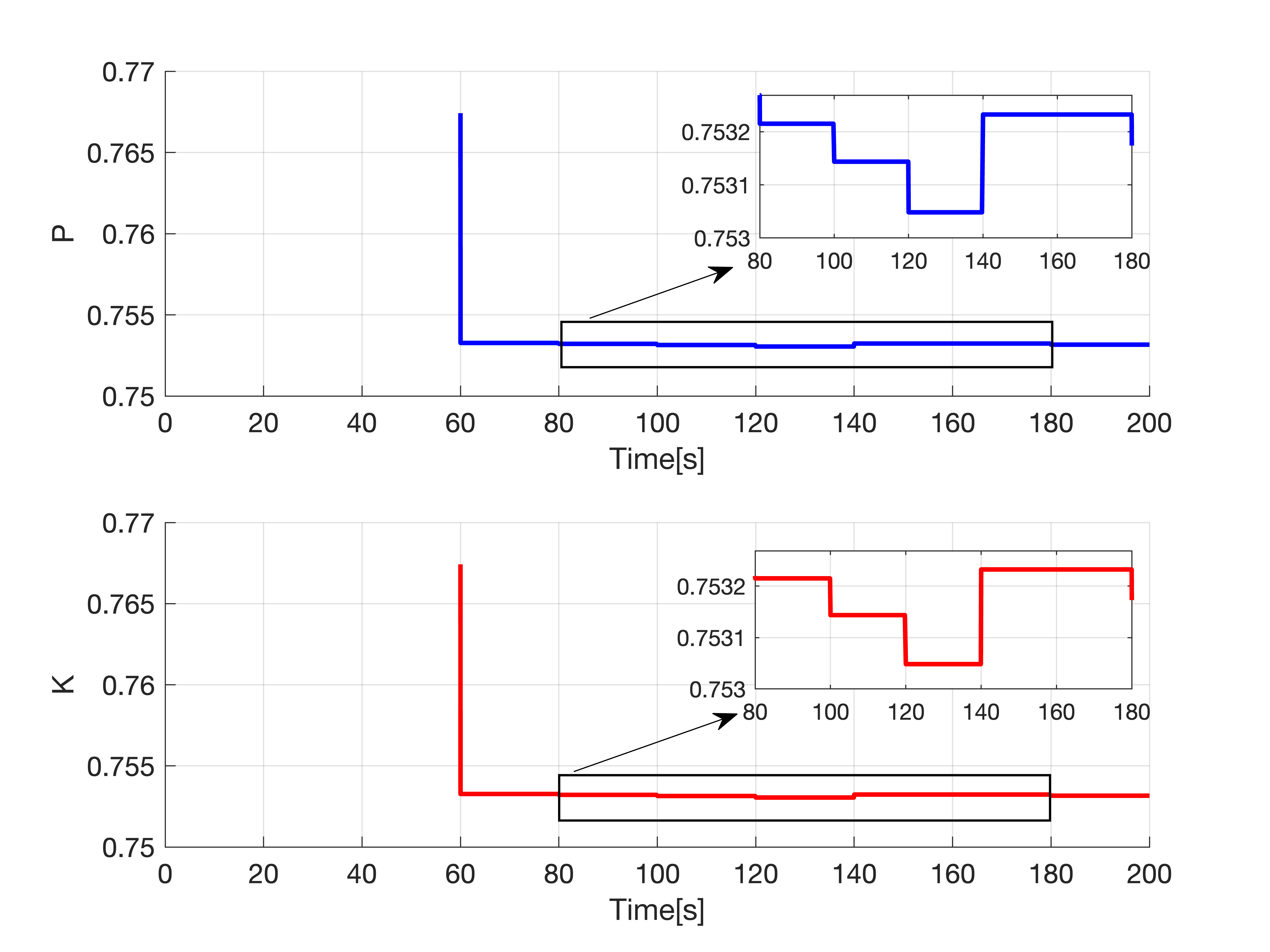}
\caption{\textit{Top Plot:} Evolution of state covariance of the proposed Kalman filtering approach for Layer 4 of \textit{Dataset \# 3} (cube geometry of size $0.4$ mm). \textit{Bottom Plot:} Evolution of Kalman gain of the proposed Kalman filtering approach for Layer 4 of \textit{Dataset \# 3} (cube geometry of size $0.4$ mm).}
\label{PK2}
\end{figure}

\begin{figure}[H]
\centering
\includegraphics[width=0.5\textwidth]{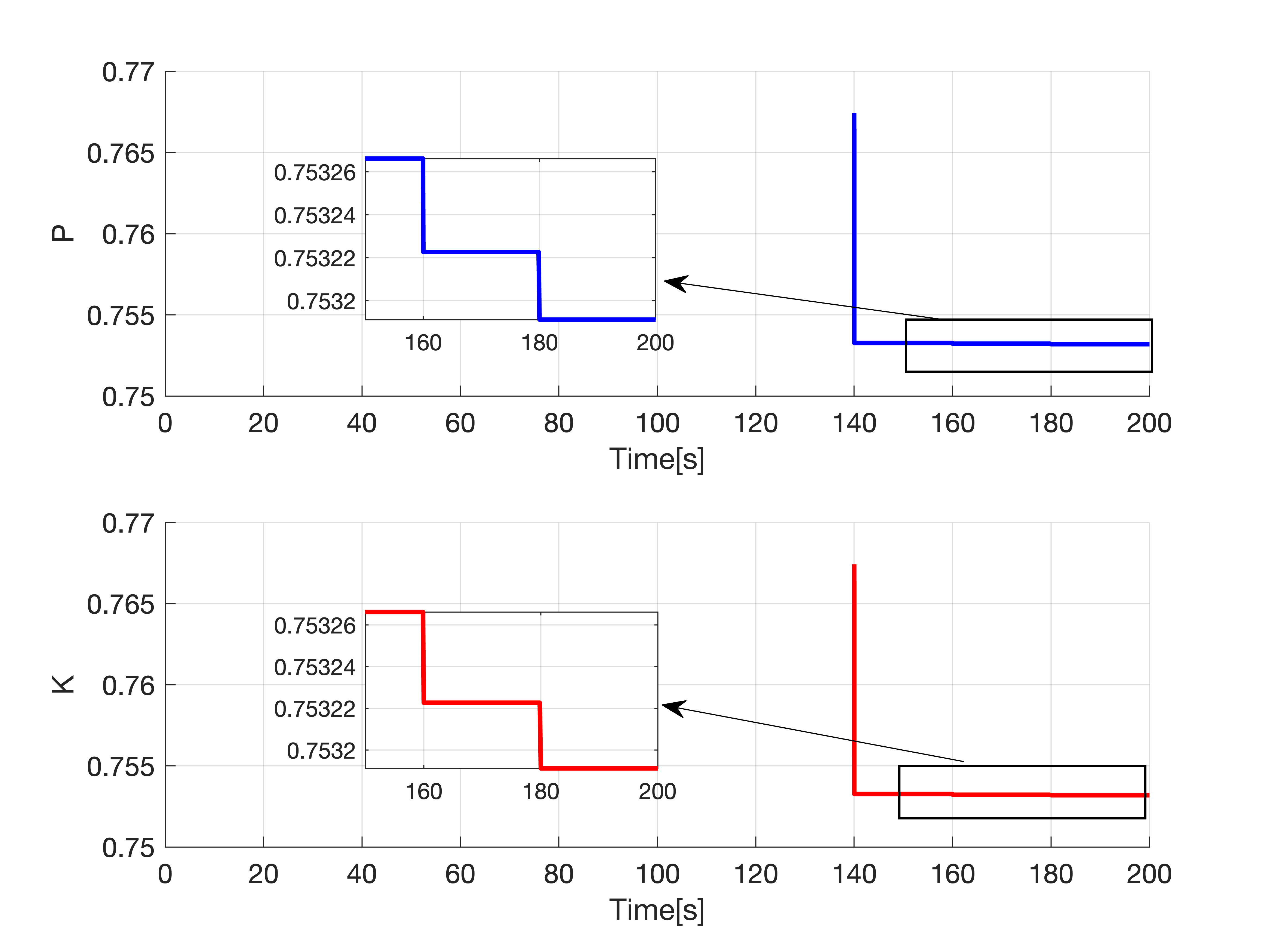}
\caption{\textit{Top Plot:}  Evolution of state covariance of the proposed Kalman filtering approach for Layer 8 of \textit{Dataset \# 3} (cube geometry of size $0.4$ mm). \textit{Bottom Plot:}  Evolution of Kalman gain of the proposed Kalman filtering approach for Layer 8 of \textit{Dataset \# 3} (cube geometry of size $0.4$ mm).}
\label{PK3}
\end{figure}

\section{Conclusions and Future Work}
This paper presented a real-time thermal state estimation and forecasting framework for laser powder bed fusion L-PBF processes, leveraging a physics-informed reduced-order model (ROM) to achieve computational efficiency while maintaining physical fidelity. The proposed framework effectively captures the spatiotemporal evolution of inter-layer thermal dynamics by incorporating governing heat transfer principles within a reduced-order setting. Model parameters were systematically identified using a genetic algorithm (GA), optimizing the reduced-order representation to align with high-fidelity thermal field data generated from ANSYS finite element simulations.

To enable robust state estimation and forecasting, the ROM was augmented with a Kalman filtering approach that assimilates real-time temperature measurements along with historical pseudo-data to reconstruct the full thermal field. This filtering strategy accounts for process noise and sensor uncertainty, enabling dynamic correction of model predictions. The framework was validated across multiple part geometries and layer configurations, demonstrating strong agreement with simulation-derived ground truth data. Notably, the proposed approach maintained high estimation accuracy even in deeper layers, where thermal accumulation and increased model uncertainty typically degrade predictive performance.

Future work will focus on transitioning the framework from simulation to practical deployment. This includes experimental validation using in-situ sensor data (e.g., pyrometers, infrared cameras, or thermographic systems) to assess the model’s real-world performance. Furthermore, the modeling approach will be extended to accommodate more complex part geometries and scan strategies, where localized heat accumulation and boundary effects become more pronounced. Integration into closed-loop control systems will also be pursued to enable real-time thermal regulation and defect mitigation during the build process. Additional efforts will target improving robustness under varying process conditions, such as changes in laser power, scan speed, or powder properties, and incorporating formal uncertainty quantification methods. These enhancements will support more reliable thermal monitoring and quality assurance in next-generation metal additive manufacturing systems.

\bibliographystyle{ieeetr}
\bibliography{ref}

\end{document}